\documentclass[aps,pre,reprint,longbibliography]{revtex4-2}
\usepackage{fontspec}
\usepackage{xeCJK}
\setCJKsansfont{HaranoAjiGothic-Regular.otf}[
  BoldFont=HaranoAjiGothic-Bold.otf]
\setCJKmonofont{HaranoAjiGothic-Regular.otf}[
  BoldFont=HaranoAjiGothic-Bold.otf]
\usepackage{amsmath,amssymb,mathtools,bm}
\usepackage{graphicx}
\usepackage{array}[=2016-10-06]
\usepackage{longtable}[=2020-01-07]
\usepackage{booktabs,array,multirow}
\usepackage{xcolor}
\usepackage{tikz}
\makeatletter
\g@addto@macro\@mkpream@relax{\let\@endpbox\relax}
\makeatother
\usepackage{hyperref}
\hypersetup{hidelinks,pdfencoding=auto}
\newcommand{\GA}{G_{\alpha}^{M_2}}

\newcommand{\avg}[1]{\left\langle #1\right\rangle}

\newcommand{\Rset}{\mathcal{U}}

\usepackage{longtable}
\begin{document}
\raggedbottom
\title{Higher-order statistical structure in perfume item combinations beyond a pairwise maximum-entropy model}
\author{Yohei Shida}
\affiliation{Institute of Systems and Information Engineering, University of Tsukuba, 1-1-1 Tennodai, Tsukuba, Ibaraki 305-8573, Japan}

\author{Jun'ichi Ozaki}
\email{ozaki.jun.fe@yokohama-cu.ac.jp}
\affiliation{School of Data Science, Yokohama City University, 22-2 Seto, Kanazawa-ku, Yokohama, Kanagawa 236-0027, Japan}
\date[]{}
\begin{abstract}
We find that deviations from a pairwise maximum-entropy model in perfume item combinations are organized around particular shared item pairs. We analyze 1,979 perfumes using 50 binary variables representing the most frequently listed items. Although the pairwise model approximately reproduces the distribution of the number of listed items per perfume, the observed three-item marginal distributions exhibit deviations exceeding those obtained from synthetic datasets subjected to the same model-refitting procedure. Strongly deviating triplets are concentrated around shared item pairs, and this concentration cannot be explained by the participation frequencies of individual items alone. We then introduce three-body couplings for 50 selected triplets and reoptimize all model parameters. The sum of squared standardized residuals over the remaining 19,550 triplets decreases by 5.44\%, with substantially larger reductions for triplets sharing an item pair with selected triplets. Furthermore, a static linear-response calculation accounting for the readjustment of lower-order parameters approximately reproduces the resulting changes in predicted co-occurrence frequencies, with a Pearson correlation coefficient of 0.850. These results reveal the organization of higher-order statistical dependencies and characterize how selected three-body couplings affect predictions throughout the system.
\end{abstract}
\maketitle

\section{Introduction}
Cultural products such as language and cuisine exhibit characteristic regularities not only in the frequencies of their constituent elements but also in how these elements are combined. In language, statistical information contained in word order has been investigated~\cite{Montemurro_2011}. For example, the frequencies of letters at individual positions and pairwise letter statistics account for approximately 92\% of the total correlation among letters in four-letter English words~\cite{Stephens_2010}. Similar regularities occur in cuisine. The tendency to combine ingredients sharing many flavor compounds differs between Western and East Asian cuisines~\cite{Ahn_2011}. Ingredient co-occurrences in recipes have also been analyzed as networks~\cite{Herrera_2021}, while changes in ingredient frequencies and combinations have been used to study culinary evolution and regional differences~\cite{Kinouchi_2008,Caprioli_2025}. These studies show that combination statistics characterize cultural products in ways not captured by the frequencies of individual elements alone.

Perfumes can similarly be studied as cultural products described by combinations of listed items. Perfume articles and databases provide item lists for individual products, allowing combinations to be compared across many perfumes. Statistical similarities among items used to describe scents have been investigated previously~\cite{Chastrette_1988}. By contrast, a study of actual perfume formulations represented combinations of raw materials as graphs and proposed a graph-search method for generating formulations~\cite{Cai_2024}. Large-scale datasets have also been used to study changes in perfumes over time. A study of perfumes released since 1900 reported that a measure of complexity based on weighted item distributions has decreased since the 1990s, whereas the novelty of item combinations has continued to increase~\cite{satarifard2026culturalevolutionperfumes1900}. That study also examined imitation relationships between products and collaborations among perfumers to investigate the cultural evolution of perfumes.

However, regularities in item combinations do not determine the order of interactions needed to describe them. Nonrandom co-occurrence patterns involving three or more items do not necessarily imply that three-body or higher-order interactions are required. Even a probabilistic model containing only pairwise interactions can exhibit higher-order correlations through indirect dependencies mediated by pairwise couplings~\cite{Schneidman_2003,Amari_2001,Morcos_2011}. For example, the collective activity of neuronal populations can, in some cases, be well described by neuronal firing rates and pairwise correlations alone~\cite{Schneidman_2006}. Nevertheless, the range of statistics accounted for by pairwise models depends on the system and conditions, and higher-order dependencies and additional statistical constraints have also been investigated~\cite{Roudi_2009,Ganmor_2011_PNAS,Tka_ik_2014,Shimazaki_2015}. Examining the co-occurrence frequencies of individual triplets alone therefore cannot distinguish higher-order correlations generated by pairwise interactions from structures that a pairwise model fails to account for. This distinction requires comparing observed three-variable marginal distributions with predictions from a model inferred from the one- and two-body statistics of the full system.

For the listed-item combinations in perfumes, it remains unclear to what extent pairwise interactions account for the observed combination structure and how the unexplained higher-order structure is organized. Deviations in three-item distributions may be scattered among individual triplets or concentrated among triplets sharing particular items or item pairs. If deviations repeatedly occur in triplets sharing the same item pair but differing in the third item, analyzing each triplet separately may obscure this common organization. Moreover, when three-body couplings are introduced for a subset of strongly deviating triplets, it is not known whether the effects remain confined to the selected combinations or also extend to triplets without added couplings. We therefore examine both the magnitude and organization of deviations from a pairwise model and the changes in predictions across the system after introducing a small number of three-body couplings.

In this study, we analyze the 50 most frequently listed items selected from 1,979 perfumes, using a pairwise maximum-entropy model inferred from one- and two-body statistics as a reference. Although this model approximately reproduces the distribution of the number of listed items per perfume, substantial deviations remain in three-item distributions. These deviations exceed the variation observed when synthetic datasets generated from the pairwise model are refitted and evaluated using the same procedure. We further find that strongly deviating triplets are concentrated around particular shared item pairs. We then introduce three-body couplings for a subset of these triplets and reoptimize all model parameters, including the one- and two-body parameters. The sum of squared residuals over the unselected triplets decreases, with a particularly large reduction for triplets sharing an item pair with a selected triplet. The corresponding changes in predicted co-occurrence frequencies are approximately reproduced by a static linear-response calculation that accounts for the readjustment of lower-order parameters.

\section{Data}
We analyzed 1,979 perfumes described in the perfume encyclopedia of Cahiers de Mode, a website featuring historically notable perfumes and products from well-known brands~\cite{CahiersDeMode_PerfumeEncyclopedia}. Each article provides information such as the perfume name, brand, and year of release, together with a list of items grouped into top, middle, and base categories. We combined these categories and recorded whether each item was listed for each perfume. The resulting data represent items listed in perfume descriptions, not measured chemical compositions or ingredient concentrations.

\begin{figure*}[!t]
 \centering
 \includegraphics[width=\textwidth]{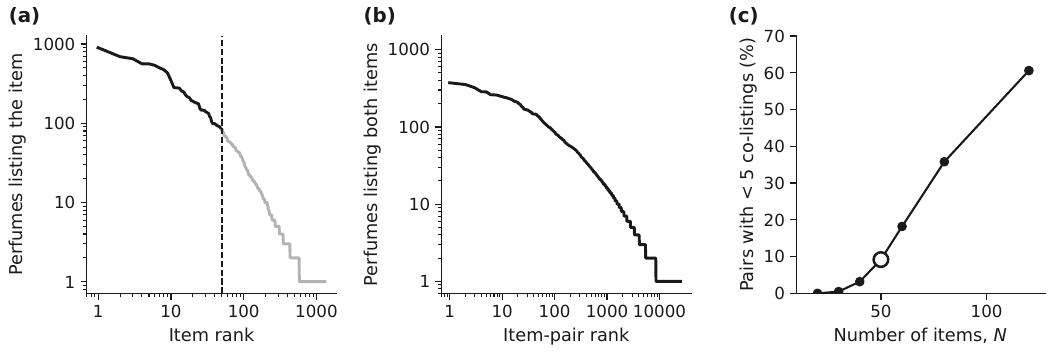}
 \caption{Listing-frequency distributions and selection of the 50 items used in the analysis. (a) Number of perfumes listing each of the 1,307 normalized items versus item rank on logarithmic axes. Black and gray curves denote the 50 most frequently listed items and the remaining items, respectively. The dashed line marks the boundary between Raspberry, ranked 50th (83 perfumes), and Pear, ranked 51st (82 perfumes). A total of 730 items (56\%) appear in only one perfume. (b) Co-occurrence frequency versus rank for distinct item pairs among all 1,307 items, shown on logarithmic axes. Only the 25,169 pairs with at least one observed co-occurrence are displayed; the 828,302 pairs with zero observed co-occurrences are omitted. (c) Fraction of item pairs with fewer than five co-occurrences among the top $N$ items ranked by listing frequency. The open circle marks $N=50$, for which 113 of 1,225 pairs (9.2\%) co-occur in fewer than five perfumes.}
 \label{fig:data}
\end{figure*}

The perfume descriptions contained 2,030 distinct item labels. We merged synonyms and spelling variants to obtain 1,307 normalized items. Figure~\ref{fig:data}(a) shows that listing frequencies are highly heterogeneous: 730 items (56\%) appear in only one perfume. Pairwise co-occurrence frequencies are also highly heterogeneous. Figure~\ref{fig:data}(b) shows the 25,169 item pairs with at least one observed co-occurrence, ranked by frequency, among all 853,471 possible pairs of the 1,307 items. The most frequent pair, Musk and Bergamot, co-occurs in 372 perfumes, whereas 16,670 observed pairs co-occur only once. A further 828,302 pairs have no observed co-occurrences.

Because these distributions are highly uneven, including less frequently listed items increases the number of pairs with few observed co-occurrences. We therefore selected the 50 most frequently listed items for analysis (Table~\ref{tab:items}). These items account for 11,712 (64.6\%) of the 18,123 item listings in the full dataset. As shown in Fig.~\ref{fig:data}(c), 113 of the 1,225 possible pairs among these 50 items (9.2\%) co-occur in fewer than five perfumes, including pairs with no observed co-occurrences. This fraction increases to 18.2\% when the 60 most frequently listed items are considered.

\begin{table*}[t]
\centering
\caption{Rank, normalized item name, and number of perfumes listing each of the 50 most frequently listed items.}
\label{tab:items}
\begin{tabular}{r l r @{\hspace{17mm}} r l r}
 \toprule
 Rank&Item&Perfumes&Rank&Item&Perfumes\\
 \midrule
 1&Musk&901&26&Grapefruit&147\\
 2&Bergamot&694&27&Lavender&147\\
 3&Jasmine&653&28&Tuberose&146\\
 4&Vanilla&564&29&Blackcurrant&145\\
 5&Patchouli&564&30&Incense&140\\
 6&Sandalwood&539&31&Peach&138\\
 7&Cedar&501&32&Geranium&135\\
 8&Rose&475&33&Cinnamon&133\\
 9&Amber&431&34&Labdanum&122\\
 10&Vetiver&348&35&Ginger&119\\
 11&Orange blossom&283&36&Coriander&108\\
 12&Mandarin orange&280&37&Orange&100\\
 13&Tonka bean&279&38&Aldehydes&99\\
 14&Iris&256&39&Clove&99\\
 15&Lemon&249&40&Peony&99\\
 16&Pink pepper&226&41&Petitgrain&97\\
 17&Neroli&215&42&Bulgarian rose&96\\
 18&Lily of the valley&210&43&Carnation&93\\
 19&Leather&194&44&Green notes&93\\
 20&Ylang-ylang&190&45&Heliotrope&92\\
 21&Violet&185&46&Nutmeg&91\\
 22&Benzoin&182&47&Magnolia&90\\
 23&Oakmoss&179&48&Saffron&87\\
 24&Cardamom&176&49&Freesia&85\\
 25&Pepper&154&50&Raspberry&83\\
 \bottomrule
\end{tabular}
\end{table*}

\section{Pairwise model and its descriptive scope}
We construct a pairwise maximum-entropy model and examine the extent to which it reproduces the one- and two-body listing frequencies used for fitting, the distribution of the number of listed items per perfume, and the three-item marginal distributions.
\subsection{Pairwise maximum-entropy model}
To investigate how far the combinations of listed items in perfumes can be described using pairwise interactions alone, we construct a pairwise maximum-entropy model with 50 binary variables. The presence or absence of each item in a perfume is represented by the binary state $\bm{x}=(x_1,\ldots,x_N)$. For perfume $a$, we set $x_i^{(a)}=1$ if item $i$ is listed and $x_i^{(a)}=0$ otherwise. The number of perfumes is $N_{\mathrm p}=1979$, and the number of items is $N=50$. The empirical listing frequencies of item $i$ and item pair $(i,j)$ are defined as

\begin{align}
 p_i&=\frac{1}{N_{\mathrm p}}\sum_{a=1}^{N_{\mathrm p}}x_i^{(a)},\nonumber\\
 p_{ij}&=\frac{1}{N_{\mathrm p}}\sum_{a=1}^{N_{\mathrm p}}x_i^{(a)}x_j^{(a)}
 \label{eq:lowmom}
\end{align}

For an arbitrary observable $U(\bm{x})$ of the listing configuration, we denote its empirical average and its expectation under a model distribution $P_\mu(\bm{x})$ by

\begin{align}
 \avg{U}_{\mathrm{data}}&=\frac{1}{N_{\mathrm p}}\sum_{a=1}^{N_{\mathrm p}}U(\bm x^{(a)}),\nonumber\\
 \avg{U}_\mu&=\sum_{\bm x\in\{0,1\}^N}U(\bm x)P_\mu(\bm x)
 \label{eq:averages}
\end{align}

We use the maximum-entropy principle to construct a probability distribution over the full listing configurations subject to constraints on the one- and two-body statistics~\cite{Jaynes_1957}. Maximizing the entropy while fixing their expectation values to the observed listing frequencies yields a distribution whose exponent contains only one- and two-body terms. This distribution is equivalent in form to an Ising model with binary variables taking values in $\{0,1\}$~\cite{Schneidman_2006}:

\begin{equation}
 P_{M_2}(\bm{x})=\frac{1}{Z_{M_2}}\exp\!\left[
 \sum_i h_i x_i+\sum_{i<j}J_{ij}x_ix_j\right]
 \label{eq:ising}
\end{equation}

Here, $h_i$ is the coefficient of $x_i$, $J_{ij}$ is the coefficient of $x_i x_j$, and $Z_{M_2}$ is the normalization constant. We refer to the model obtained by regularized maximum-likelihood estimation within this distribution family as the pairwise model $M_2$.

We estimate the parameters by maximizing the penalized average log-likelihood with $L_2$ regularization on the pairwise couplings $J_{ij}$~\cite{Ekeberg_2013}:

\begin{equation}
 \mathcal L_{M_2}=\frac{1}{N_{\mathrm p}}\sum_{a=1}^{N_{\mathrm p}}\log P_{M_2}(\bm{x}^{(a)})
 -\frac{\gamma_J}{2}\sum_{i<j}J_{ij}^2
 \label{eq:likelihood}
\end{equation}

Here, $\gamma_J$ is the regularization coefficient for the pairwise couplings. The stationarity conditions are $\avg{x_i}_{M_2}=p_i$ and $\avg{x_ix_j}_{M_2}=p_{ij}-\gamma_JJ_{ij}$. Thus, under regularized estimation, the model expectations of the pairwise listing frequencies do not generally coincide exactly with their empirical values. We evaluate model expectations using Gibbs sampling~\cite{Geman_1984} and iteratively update the parameters.

\subsection{Reproduction of low-order statistics and the item-count distribution}
We first compare the empirical one- and two-body listing frequencies used in fitting with the corresponding model predictions.

We assessed deviations in the one- and two-body listing frequencies using dimensionless descriptive standardized differences. For the model predictions $p_i^{M_2}$ and $p_{ij}^{M_2}$, we used the listing frequencies averaged over the 200 synthetic datasets generated with the model parameters held fixed. The differences between the empirical and model values are defined as

\begin{align}
 z_i&=\frac{p_i-p_i^{M_2}}{\sqrt{p_i(1-p_i)/N_{\mathrm p}}},\nonumber\\
 z_{ij}&=\frac{p_{ij}-p_{ij}^{M_2}}{\sqrt{p_{ij}(1-p_{ij})/N_{\mathrm p}}}
 \label{eq:marginal_z_main}
\end{align}

The denominators are reference standard deviations for listing frequencies under a binomial sampling model. The standardized difference is undefined when the corresponding empirical frequency $p$ is 0 or 1, because the denominator vanishes.

\begin{figure*}[!t]
 \centering
 \includegraphics[width=0.94\textwidth]{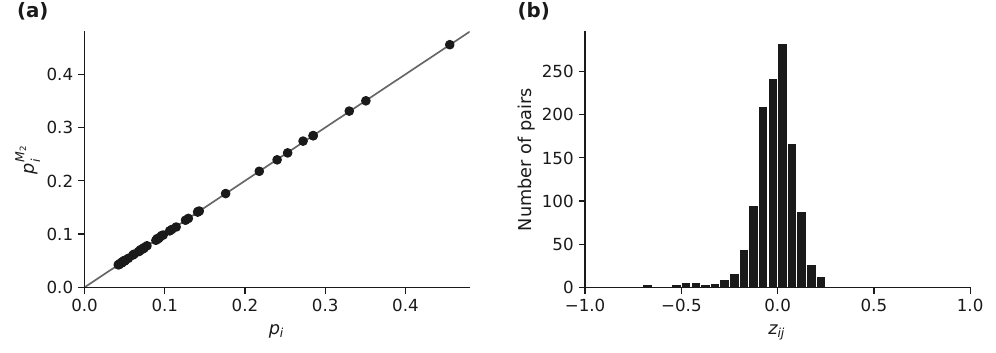}
 \caption{Reproduction of low-order moments by the pairwise model. Model predictions are listing frequencies averaged over 200 synthetic datasets, each corresponding to 1,979 perfumes. (a) Empirical one-body listing frequencies $p_i$ and model predictions $p_i^{M_2}$. The gray line indicates equality. (b) Distribution of the standardized differences $z_{ij}$ in pairwise listing frequencies defined in Eq.~(\ref{eq:marginal_z_main}). The 1,217 item pairs with nonzero empirical co-occurrence are shown; the eight pairs with zero observed co-occurrence are omitted. Differences are computed as the empirical value minus the model prediction, and the differences in pairwise moments include contributions from $L_2$ regularization. The root-mean-square standardized differences are 0.079 for the 50 one-body statistics and 0.112 for the 1,217 nonzero-co-occurrence item pairs; the corresponding maximum absolute values are 0.22 and 0.72.}
 \label{fig:s_marginals}
\end{figure*}

Figure~\ref{fig:s_marginals} shows close agreement between the fitted pairwise model and the empirical lower-order listing frequencies. The root-mean-square standardized differences are 0.079 for the 50 one-body statistics and 0.112 for the 1,217 item pairs with nonzero observed co-occurrences. The corresponding maximum absolute values are 0.22 and 0.72.

The standardized difference is undefined for eight of the 1,225 item pairs because their observed co-occurrence frequencies are zero. These eight pairs are excluded from the pairwise RMS and from Fig.~\ref{fig:s_marginals}(b). Their pairwise-model predicted co-occurrence counts, averaged over the same 200 fixed-parameter synthetic datasets, are all below one (0.53 to 0.79 perfumes); the RMS difference from the observed zero counts is 0.686 perfumes.

We next examine how well the pairwise model reproduces the distribution of the number of listed items per perfume. We define $K=\sum_i x_i$ as the number of listed items among the 50 selected items. The one- and two-body listing frequencies determine the mean and variance of $K$, but do not uniquely determine its full distribution. We therefore compare the observed distribution with that obtained from synthetic datasets generated by the pairwise model.

\begin{figure*}[!t]
 \centering
 \includegraphics[width=\textwidth]{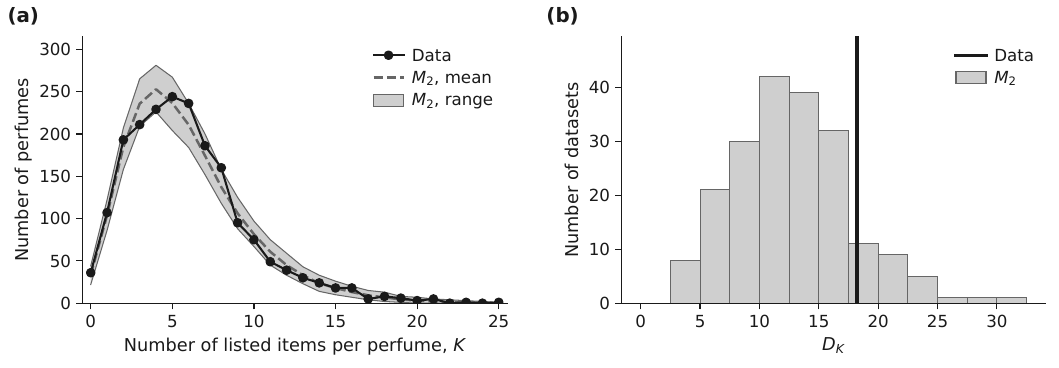}
 \caption{Reproduction of the distribution of the number of listed items per perfume, $K$, by the pairwise model. Here, $K$ is the number of listed items among the 50 selected items. (a) Black points and lines show the observed distribution. The gray dashed line and shaded band indicate the mean and the 2.5th to 97.5th percentiles, respectively, at each $K$ across 200 synthetic datasets generated from the pairwise model, each containing 1,979 perfumes. (b) Histogram of $D_K$ for the 200 synthetic datasets, compared with the observed value. The distance is calculated using 14 bins: $K=0,\ldots,12$ and $K\geq13$. Gray bars show the synthetic-dataset distribution, and the black vertical line indicates the observed value, $D_K=18.21$. The mean for the synthetic datasets is $12.78$.}
 \label{fig:k}
\end{figure*}

The pairwise model approximately reproduces the observed item-count distribution. Keeping the fitted model parameters fixed, we generate 200 synthetic datasets, each containing 1,979 configurations of the 50 binary variables. As shown in Fig.~\ref{fig:k}(a), the observed distribution agrees approximately with the distributions obtained from the synthetic datasets. To quantify this agreement, we define a distance $D_K$ using the observed count $n_b$ in bin $b$ and the mean count $\overline n_b^{M_2}$ over the 200 synthetic datasets:

\begin{equation}
 D_K=\sum_{b:\,\overline n_b^{M_2}>0}
 \frac{(n_b-\overline n_b^{M_2})^2}{\overline n_b^{M_2}}
 \label{eq:DK_main}
\end{equation}

As shown in Fig.~\ref{fig:k}(b), the observed value $D_K=18.21$ lies within the 2.5th to 97.5th percentile interval of the synthetic datasets, $4.20$ to $23.54$. The mean value for the synthetic datasets is $12.78$.

\subsection{Deviations in three-item distributions}
To what extent can the pairwise model account for the statistical structure of three-item combinations? To address this question, we calculate the three-variable marginal distributions of the full 50-variable pairwise model and compare them with the empirical distributions. We denote a triplet of distinct items by $\alpha=\{i,j,k\}$, where $i<j<k$, and the set of all triplets by $\mathcal{T}$. The 50 selected items yield 19,600 distinct triplets. Let $\bm{s}=(s_i,s_j,s_k)\in\{0,1\}^3$ denote a three-item listing state, and let $n_\alpha(\bm{s})$ be its observed count. Writing $\bm{x}_\alpha=(x_i,x_j,x_k)$ and denoting the remaining 47 items by $\bar{\alpha}$, the marginal probability predicted by the pairwise model is

\begin{equation}
 q_\alpha^{M_2}(\bm s)
 =\sum_{\bm x_{\bar\alpha}\in\{0,1\}^{N-3}}
 P_{M_2}(\bm x_\alpha=\bm s,\bm x_{\bar\alpha})
 \label{eq:marginal}
\end{equation}

To quantify the discrepancy between the empirical distribution and the pairwise-model prediction, we use a likelihood-ratio-type deviance based on the observed counts of all eight states~\cite{Cressie_1984}:

\begin{equation}
 \GA=2\sum_{\bm{s}\in\{0,1\}^3}
 n_\alpha(\bm{s})
 \log\frac{n_\alpha(\bm{s})}{N_{\mathrm p}q_\alpha^{M_2}(\bm{s})}
 \label{eq:GA}
\end{equation}

Terms with $n_\alpha(\bm{s})=0$ are taken to be zero. A larger $G_\alpha^{M_2}$ indicates a greater discrepancy in the three-variable marginal distribution. To distinguish the direction of the deviation, we focus on the state $(1,1,1)$, in which all three items are listed. We define its observed count as $o_\alpha=n_\alpha(1,1,1)$ and its predicted count under the pairwise model as $e_\alpha^{M_2}=N_{\mathrm p}q_\alpha^{M_2}(1,1,1)$. A triplet has a positive deviation when $o_\alpha>e_\alpha^{M_2}$ and a negative deviation when $o_\alpha<e_\alpha^{M_2}$. Throughout the figures, orange and blue indicate positive and negative deviations, respectively; the nonpositive group in Fig.~\ref{fig:s_threshold}(a) also includes equality.

We compare the observed deviations in the three-item distributions with those obtained from synthetic datasets generated by the pairwise model. We generate 200 datasets of 1,979 configurations each from the fitted pairwise model and refit a 50-variable pairwise model to each dataset. We retain the 194 datasets that satisfy the fitting termination criterion and evaluate the triplet deviances using the same procedure as for the observed data.

\begin{figure*}[!t]
 \centering
 \includegraphics[width=0.89\textwidth]{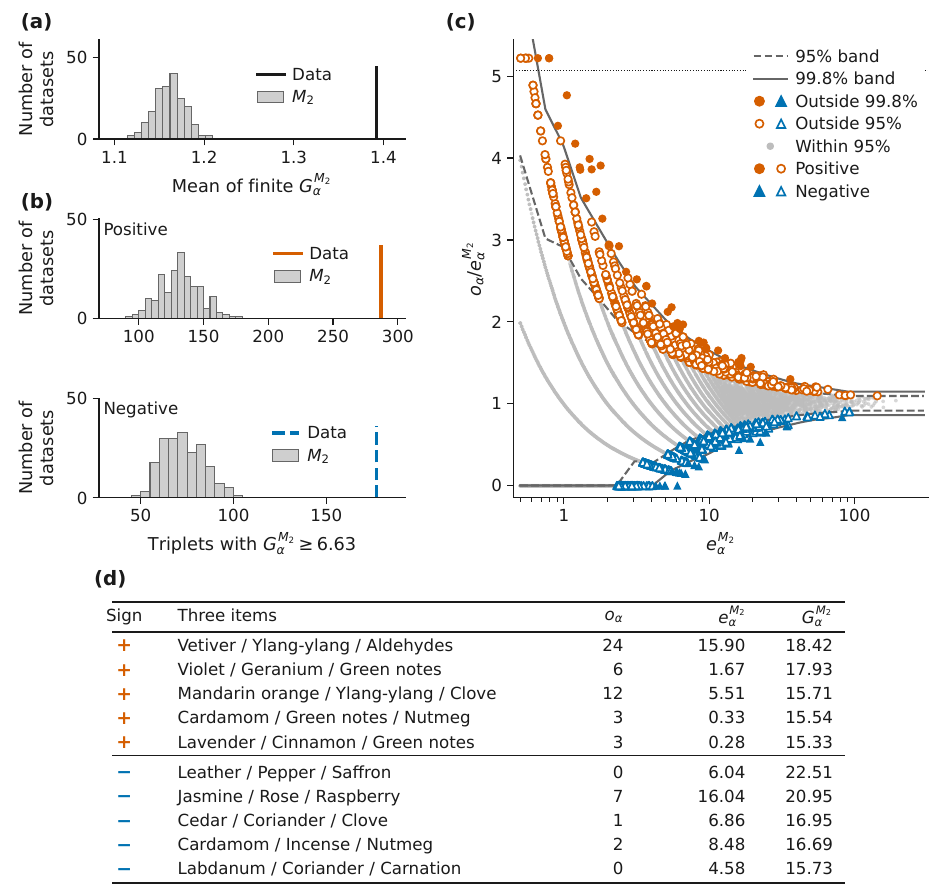}
 \caption{Deviations from the three-variable marginal distributions predicted by the pairwise model. (a) The black vertical line indicates the mean $G_\alpha^{M_2}$ over all 19,600 triplets in the observed data. Gray bars show the distribution of the mean finite deviance across 194 synthetic datasets for which the pairwise model was refitted. (b) Numbers of positive and negative triplets satisfying $G_\alpha^{M_2}\geq6.63$. Gray bars show the corresponding distributions from the same synthetic datasets. The orange solid and blue dashed lines indicate the observed positive and negative counts, respectively. Triplets assigned $G_\alpha^{M_2}=+\infty$ numerically because of zero Monte Carlo probability estimates are excluded from the averages in (a) but included in the above-threshold counts in (b). Numerical details are given in Appendix~\ref{sec:app_GA}. (c) Funnel plot of the observed-to-predicted ratio $o_\alpha/e_\alpha^{M_2}$ (vertical axis) against the predicted three-item co-occurrence count $e_\alpha^{M_2}$ (horizontal axis). The plot includes 16,513 triplets with $e_\alpha^{M_2}\geq0.5$. Dashed and solid curves indicate the 95\% and 99.8\% empirical bands constructed from synthetic datasets, respectively. Orange circles and blue triangles indicate positive and negative deviations. Filled symbols lie outside the 99.8\% band, open symbols lie outside the 95\% band but within the 99.8\% band, and light gray points lie within the 95\% band. The nine points with ratios exceeding 5 are displayed above the upper dotted line. (d) Five triplets with large $G_\alpha^{M_2}$ for each deviation sign. Here, $o_\alpha$ denotes the observed three-item co-occurrence count.}
 \label{fig:GA}
\end{figure*}

Figure~\ref{fig:GA}(a) shows that the mean deviance over all 19,600 triplets is $1.392$ for the observed data, compared with an average of $1.161$ across the 194 synthetic datasets. The observed mean exceeds that of every refitted synthetic dataset and lies well beyond the upper end of their distribution. Thus, the deviations in the observed three-item distributions cannot be accounted for by the sample-to-sample variation obtained through generation and refitting under the pairwise model alone.

Using the upper 1\% point of a chi-squared distribution with one degree of freedom as a reference, we set a common comparison threshold of $\tau=6.63$. As shown in Fig.~\ref{fig:GA}(b), 464 triplets exceed this threshold in the observed data, including 287 positive and 177 negative deviations. This is approximately 2.3 times the mean count of $203.9$ in the 194 synthetic datasets. None of the synthetic datasets produces as many above-threshold triplets as the observed data. The discrepancy seen in Fig.~\ref{fig:GA}(a) is therefore also evident in the number of strongly deviating triplets. The excess persists over different comparison thresholds (Appendix~\ref{sec:app_GA}).

Three-item co-occurrence counts exhibit deviations in both directions relative to the pairwise-model predictions. Figure~\ref{fig:GA}(c) plots the observed-to-predicted ratio $o_\alpha/e_\alpha^{M_2}$ against the predicted count $e_\alpha^{M_2}$. The observed ratios extend well beyond both the upper and lower boundaries of the 95\% and 99.8\% empirical bands obtained from synthetic datasets, revealing substantial deviations in individual triplets.

For example, the triplet Vetiver, Ylang-ylang, and Aldehydes in Fig.~\ref{fig:GA}(d) has an observed count of 24, compared with the pairwise-model prediction of $15.90$, giving $G_\alpha^{M_2}=18.42$. In contrast, Leather, Pepper, and Saffron has no observed co-occurrences, despite a predicted count of $6.04$, giving $G_\alpha^{M_2}=22.51$. These examples demonstrate both combinations that occur substantially more often than predicted and combinations whose predicted co-occurrences are absent from the data. Both deviances are well above the reference threshold of $6.63$. Deviations in the three-item distributions also remain under local pairwise models that preserve the one- and two-body frequencies of each triplet (Appendix~\ref{sec:app_localG}).

Pairwise generative models with item-specific fields varying across release-year or brand groups also fail to reproduce the observed deviations. The deviations persist after combining similar products within the same brand. These results indicate that neither heterogeneity across groups nor duplication of similar products alone accounts for the deviations in the three-item distributions (Appendix~\ref{sec:app_heterogeneity}).

\section{Shared-pair structure of deviating triplets}
We next examine whether the triplets deviating from the pairwise model exhibit a common structure. In particular, we focus on the concentration of deviating triplets around shared item pairs. We divide triplets satisfying $G_\alpha^{M_2}\geq 6.63$ into the positive and negative sets $\mathcal D^+$ and $\mathcal D^-$ according to whether the observed three-item co-occurrence count is larger or smaller than the model prediction. The observed data contain 287 positive and 177 negative triplets. We define the number of deviating triplets containing item pair $(i,j)$ as

\begin{equation}
 c_{ij}^{\pm}=\sum_{\alpha\in\mathcal D^{\pm}}
 \mathbf{1}\bigl[\{i,j\}\subset\alpha\bigr]
 \label{eq:cij_main}
\end{equation}

Here, $\mathbf{1}[\cdot]$ denotes the indicator function. We further define the total number of pairs of distinct triplets sharing an item pair as

\begin{equation}
 T^{\pm}=\sum_{i<j}\binom{c_{ij}^{\pm}}{2}
 \label{eq:T}
\end{equation}

Because $T^\pm$ also depends on the number of selected triplets, we match the selection counts when comparing with synthetic datasets. For each of the 194 synthetic datasets for which the pairwise model was successfully refitted, we select the 287 positive and 177 negative triplets with the largest $G_\alpha^{M_2}$ within each sign.

To distinguish shared-pair concentration from the effect of individual items appearing in many deviating triplets, we randomize the triplet sets while preserving the number of occurrences of each item. The numbers of occurrences of individual item pairs are not held fixed. This control tests whether the concentration around shared item pairs could arise solely from individual-item participation frequencies and is related to the configuration-model approach for higher-order networks~\cite{Chodrow_2020}. We use 2,000 randomizations of the observed deviating-triplet sets; for plotting, the histogram counts are rescaled to correspond to 194 datasets.

Denoting the mean $T^\pm$ over randomized sets by $\langle T^\pm\rangle_{\mathrm{rand}}$, we define the excess shared-pair concentration as

\begin{equation}
 X^\pm=T^\pm-\avg{T^\pm}_{\mathrm{rand}}
 \label{eq:excess_main}
\end{equation}

\begin{figure*}[!t]
 \centering
 \includegraphics[width=0.95\textwidth]{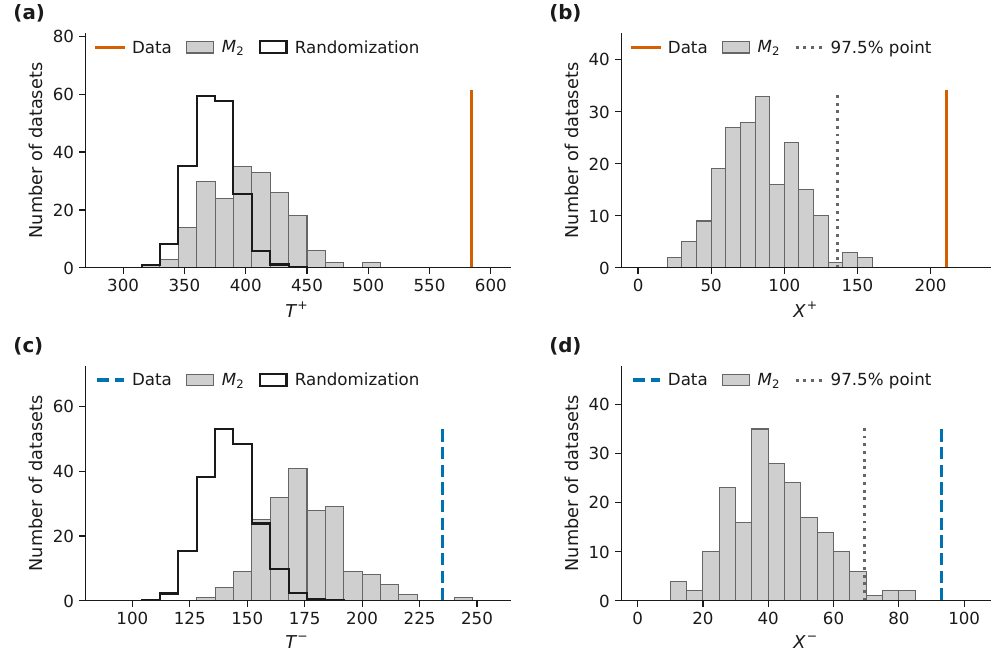}
 \caption{Shared-pair concentration of deviating triplets and randomization preserving item participation counts. (a) and (c) Shared-pair concentration $T^+$ for 287 positive triplets and $T^-$ for 177 negative triplets, respectively. Gray bars show distributions from 194 refitted synthetic datasets, with equal numbers of deviating triplets selected for each sign. Black step curves show the distributions from 2,000 randomizations of the observed deviating-triplet sets preserving the participation count of each item, with histogram counts rescaled to correspond to 194 datasets. (b) and (d) Excess concentration $X^+$ and $X^-$ relative to the randomization mean, respectively. Gray histograms and dotted vertical lines represent the distributions of $X^\pm$ and their 97.5th percentiles obtained by applying the randomization to each synthetic dataset. Orange solid and blue dashed vertical lines indicate the observed positive and negative values, respectively.}
 \label{fig:s_sharing}
\end{figure*}

We apply this procedure to the deviating-triplet sets from the observed data and from each synthetic dataset, and compare the resulting $T^\pm$ and $X^\pm$ distributions in Fig.~\ref{fig:s_sharing}.

Figure~\ref{fig:s_sharing} shows shared-pair concentrations of $T^+=584$ and $T^-=235$ for the observed positive and negative deviations in panels (a) and (c), respectively. These values exceed the corresponding means of 400.3 and 174.2 across the 194 refitted synthetic datasets and lie outside their respective 95\% intervals of 347-459 and 144-211.

The randomization preserving individual-item participation counts does not account for the observed shared-pair concentration. The observed excesses are $X^+=211.0$ and $X^-=93.1$, exceeding the corresponding synthetic-dataset means of 83.6 and 42.3, as shown in Fig.~\ref{fig:s_sharing}(b) and (d). Both observed excesses lie on the high side of their respective synthetic-dataset distributions. Thus, deviations are concentrated among triplets sharing the same item pair, beyond the effect of particular items repeatedly appearing in deviating triplets.

\begin{figure}[tbp]
 \centering
 \includegraphics[width=\columnwidth]{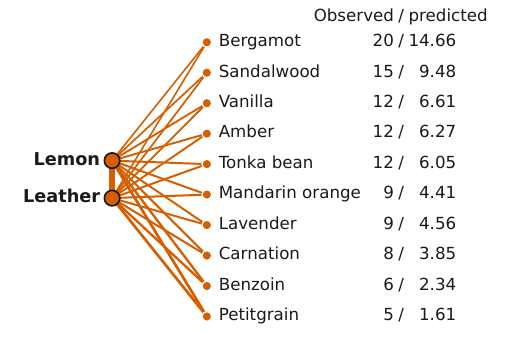}
 \caption{Ten triplets with positive deviations that share Lemon and Leather. For each third item, the two numbers give the observed three-item co-occurrence count $o_\alpha$ and the pairwise-model prediction $e_\alpha^{M_2}$, respectively. Edge widths represent the observed-to-predicted ratios $o_\alpha/e_\alpha^{M_2}$.}
 \label{fig:sharing}
\end{figure}

For example, Fig.~\ref{fig:sharing} shows ten triplets with positive deviations that share Lemon and Leather but differ in the third item.

\section{Three-body couplings and system-wide response}
\subsection{Three-body couplings and residual changes}
The deviations in the three-item distributions are concentrated among triplets sharing particular item pairs. We therefore introduce three-body couplings for a subset of strongly deviating triplets and investigate how predictions change for other triplets without added couplings. There are ten triplets with pairwise-model deviance $G_\alpha^{M_2}\geq15.14$, and 62 with $G_\alpha^{M_2}\geq10.83$. In view of these numbers, we introduce three-body couplings incrementally for the top $m=10,20,30,40,$ and $50$ triplets ranked by deviance. Let $S_m$ denote the set of the top $m$ triplets. For triplet $\alpha=\{i,j,k\}$, we define the three-body statistic $\psi_\alpha(\bm{x})=x_i x_j x_k$ and its coupling $\kappa_\alpha$. The three-body model $M_{3,m}$ is then defined as

\begin{equation}
 \begin{aligned}
 P_{M_{3,m}}(\bm{x})=\frac{1}{Z_{M_{3,m}}}\exp\Bigl[&
 \sum_i h_i x_i+\sum_{i<j}J_{ij}x_ix_j\\
 &+\sum_{\alpha\in S_m}
 \kappa_\alpha\psi_\alpha(\bm{x})\Bigr]
 \end{aligned}
 \label{eq:tripletmodel}
\end{equation}

Here, $Z_{M_{3,m}}$ is the partition function. We use the same symbols $h_i$ and $J_{ij}$ as in the pairwise model, but their fitted values depend on the model.

We reoptimize the one-body parameters $h_i$ and pairwise couplings $J_{ij}$ together with the selected three-body couplings. We define the penalized average log-likelihood with $L_2$ regularization on both the pairwise and three-body couplings as

\begin{align}
 \mathcal L_{M_{3,m}}={}&\frac1{N_{\mathrm p}}\sum_{a=1}^{N_{\mathrm p}}
 \log P_{M_{3,m}}(\bm{x}^{(a)})
 -\frac{\gamma_J}{2}\sum_{i<j}J_{ij}^2\nonumber\\
 &-\frac{\gamma_\kappa}{2}\sum_{\alpha\in S_m}\kappa_\alpha^2
 \label{eq:LG_main}
\end{align}

Here, $\gamma_\kappa$ denotes the regularization coefficient for the three-body couplings $\kappa_\alpha$. Writing $o_\alpha$ for the observed three-item co-occurrence count of a selected triplet, the stationarity condition for its coupling is

\begin{equation}
 \frac{o_\alpha}{N_{\mathrm p}}
 -\avg{\psi_\alpha}_{M_{3,m}}
 -\gamma_\kappa\kappa_\alpha=0
 \quad(\alpha\in S_m)
 \label{eq:kap_station_main}
\end{equation}

All parameters are refitted separately for each $m$.

To compare the effects of adding three-body couplings on unselected triplets, we use a common evaluation set that does not depend on $m$. We exclude the 50 selected triplets $S_{50}$ from the set $\mathcal T$ of all triplets and define

\begin{equation}
 \Rset=\mathcal T\setminus S_{50},
 \qquad|\Rset|=19\,550
 \label{eq:fixedU}
\end{equation}

Fixing this set for every model $M_{3,m}$ allows us to compare residuals over the same 19,550 triplets as the number of added couplings changes. No explicit three-body couplings are assigned to these triplets, but their three-variable marginal distributions may change because all model parameters are reoptimized.

Let $o_\beta$ be the observed three-item co-occurrence count for triplet $\beta$, and let $\widehat t_\beta^\mu$ be its co-occurrence probability estimated from an independent Monte Carlo (MC) evaluation sample of model $\mu\in\{M_2,M_{3,m}\}$. Using the pairwise-model reference probability $\widetilde t_\beta^{M_2}$ in a denominator common to all models, we define the standardized residual for each evaluation sample as

\begin{equation}
 z_\beta^\mu=
 \frac{o_\beta-N_{\mathrm p}\widehat t_\beta^\mu}
 {\sqrt{N_{\mathrm p}\widetilde t_\beta^{M_2}(1-\widetilde t_\beta^{M_2})}}
 \label{eq:z_main}
\end{equation}

With $\langle\cdot\rangle_{\mathrm{MC}}$ denoting the average over independent evaluation samples, the sum of squared residuals over a set $X$ is

\begin{equation}
 E_X(\mu)=\sum_{\beta\in X}
 \left\langle (z_\beta^\mu)^2\right\rangle_{\mathrm{MC}}
 \label{eq:E_main}
\end{equation}

The residual reduction percentage over the fixed evaluation set $\Rset$ is defined as

\begin{equation}
 R_{\Rset}(m)=100\left[
 1-\frac{E_{\Rset}(M_{3,m})}{E_{\Rset}(M_2)}
 \right]\ (\%)
 \label{eq:reduction}
\end{equation}

We select the triplets to which couplings are added using $G_\alpha^{M_2}$, calculated from all eight states of the three-variable marginal distribution. By contrast, the residual evaluation is based on the observed and predicted counts for the three-item co-occurrence state $(1,1,1)$.

\begin{figure*}[!t]
 \centering
 \includegraphics[width=0.94\textwidth]{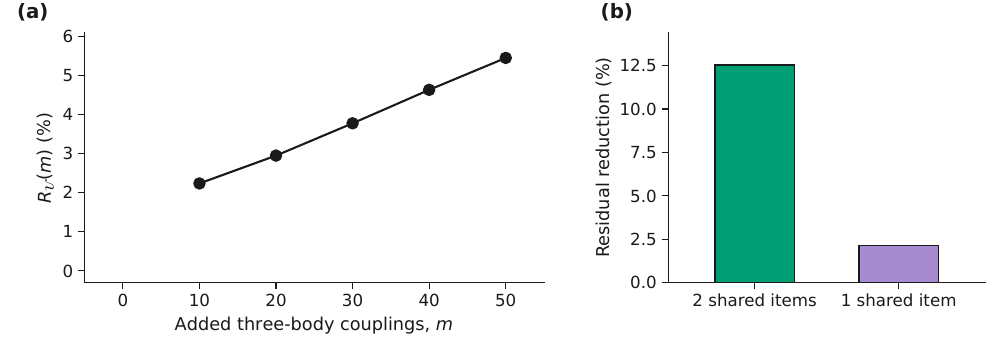}
 \caption{Changes in residuals for unselected triplets following the introduction of three-body couplings. (a) Residual reduction $R_{\Rset}(m)$ over the fixed evaluation set $\Rset=\mathcal T\setminus S_{50}$ of 19,550 triplets after adding three-body couplings to the top $m=10,20,30,40,$ and $50$ triplets and reoptimizing all parameters. Here, $\mathcal T$ is the set of all triplets and $S_{50}$ is the set of 50 selected triplets; the evaluation set is identical for all models. Points indicate the mean residual reduction over four independent MC evaluation samples for each three-body model. (b) Residual reductions for $m=50$ grouped by the maximum number of shared items $\ell(\beta)$ with the selected triplets. Green and purple bars show 5,553 triplets sharing two items and 13,977 triplets sharing one item, respectively. Each reduction is calculated relative to the pairwise-model sum of squared standardized residuals within that group.}
 \label{fig:improve}
\end{figure*}

Introducing three-body couplings also reduces the sum of squared residuals for triplets without explicitly added couplings. Figure~\ref{fig:improve}(a) shows residual reductions of 2.23\%, 2.94\%, 3.77\%, 4.63\%, and 5.44\% over the fixed evaluation set of 19,550 triplets for $m=10,20,30,40,$ and $50$, respectively. The increasing reduction with the number of couplings indicates that introducing three-body couplings for a subset of triplets, together with reoptimization of all parameters, affects predictions for many other triplets without added couplings.

The reduction depends on the number of items shared between the unselected and selected triplets. To quantify this relationship, we define the maximum number of shared items between an unselected triplet $\beta$ and the selected set $S_{50}$ as

\begin{equation}
 \ell(\beta)=\max_{\alpha\in S_{50}}
 |\alpha\cap\beta|
 \label{eq:ell_main}
\end{equation}

We use green and purple for overlap classes involving two and one shared items, respectively, and gray for zero shared items where applicable. In Fig.~\ref{fig:improve}(b), the classes are defined by the maximum overlap $\ell(\beta)$ with the selected set. As shown in Fig.~\ref{fig:improve}(b), the residual reduction is 12.53\% for the 5,553 unselected triplets sharing two items with a selected triplet, compared with 2.15\% for the 13,977 triplets sharing one item. Each percentage is calculated relative to the sum of squared standardized residuals under the pairwise model within the corresponding group. The larger improvement for triplets sharing an item pair indicates that changes in predictions induced by the three-body couplings depend on the overlap between triplets.

\subsection{Effective linear response}
We next examine whether the changes in predictions caused by adding three-body couplings can be accounted for by covariances of the original pairwise model. Since the one- and two-body parameters are also reoptimized when three-body couplings are introduced, we consider a linear response that incorporates the changes in these lower-order parameters. Let $\bm\phi(\bm{x})$ be the vector of one-body statistics $x_i$ and pairwise statistics $x_i x_j$, with corresponding parameter vector $\bm\theta$. For a selected set $S$ of $m=|S|$ triplets, let $\bm\psi_S$ be the vector of three-body statistics $\psi_\alpha(\bm{x})=x_i x_j x_k$ for $\alpha=\{i,j,k\}$, and let $\bm\kappa_S$ be the corresponding coupling vector.

For a Gibbs distribution, the change in the expectation of an observable with respect to a parameter is given by its covariance with the statistic coupled to that parameter~\cite{Kubo_1957,Kubo_1966,Kappen_1998}. Specifically, for statistics $U$ and $V$ independent of $\lambda$, where $\lambda$ multiplies $V$ in the exponent, we have

\begin{equation}
 \frac{\partial\avg{U}}{\partial\lambda}
 =\operatorname{Cov}(U,V)
 \label{eq:s_cov}
\end{equation}

We define the pairwise-model covariance $\operatorname{Cov}_{M_2}(U,V)=\langle UV\rangle_{M_2}-\langle U\rangle_{M_2}\langle V\rangle_{M_2}$ and the covariance blocks

\begin{align}
 F&=\operatorname{Cov}_{M_2}(\bm\phi,\bm\phi),\nonumber\\
 B&=\operatorname{Cov}_{M_2}(\bm\phi,\bm\psi_S),\nonumber\\
 C&=\operatorname{Cov}_{M_2}(\bm\psi_S,\bm\psi_S)
 \label{eq:blocks_main}
\end{align}

To incorporate the reoptimization of lower-order parameters, we linearize the regularized stationarity conditions around the pairwise model. Let $\Gamma$ be the diagonal matrix with zero entries for one-body parameters and $\gamma_J$ for pairwise couplings. Denote the pairwise-model parameter vector by $\bm\theta_{M_2}$, and write $\delta\bm\theta=\bm\theta-\bm\theta_{M_2}$ and $\delta\bm\kappa_S=\bm\kappa_S$. Let $\widehat{\bm\psi}_S$ be the vector of observed means of the selected three-body statistics, and define their residual relative to the pairwise model by $\bm r_S=\widehat{\bm\psi}_S-\langle\bm\psi_S\rangle_{M_2}$. The stationarity conditions to first order then give

\begin{equation}
 \begin{pmatrix}
 F+\Gamma&B\\
 B^\mathsf T&C+\gamma_\kappa I_m
 \end{pmatrix}
 \begin{pmatrix}\delta\bm\theta\\\delta\bm\kappa_S\end{pmatrix}
 \simeq
 \begin{pmatrix}\bm 0\\\bm r_S\end{pmatrix}
 \label{eq:s_joint_linear_system}
\end{equation}

Here, $I_m$ is the $m$-dimensional identity matrix. A detailed derivation of the stationarity conditions is given in Appendix~\ref{sec:app_response}.

Eliminating the changes in lower-order parameters yields the effective response matrix for the three-body couplings,

\begin{equation}
 \Sigma^\perp_{SS}=
 C-B^\mathsf T(F+\Gamma)^{-1}B
 \label{eq:sigmaSS_main}
\end{equation}

The second term accounts for the readjustment of lower-order parameters, and $\Sigma^\perp_{SS}$ is the Schur complement including regularization.

Let $t_\beta=\langle\psi_\beta\rangle$ be the three-item co-occurrence probability for an unselected triplet $\beta\notin S$. We define its effective response to the selected set $S$ as

\begin{align}
 \Sigma^\perp_{\beta S}={}&
 \operatorname{Cov}_{M_2}(\psi_\beta,\bm\psi_S)\nonumber\\
 &-\operatorname{Cov}_{M_2}(\psi_\beta,\bm\phi)
 (F+\Gamma)^{-1}B
 \label{eq:sigma}
\end{align}

The change in its predicted co-occurrence probability is then approximated by

\begin{equation}
 \delta t_\beta^{\mathrm{LR}}\simeq
 \Sigma^\perp_{\beta S}
 (\Sigma^\perp_{SS}+\gamma_\kappa I_m)^{-1}\bm r_S
 \label{eq:s_closed_response}
\end{equation}

The right-hand side depends only on the observed residuals of the selected triplets, covariances of the pairwise model, and the regularization coefficients. Thus, the first-order response of unselected triplets can be calculated without using the fitted parameters or predictions of the three-body model.

\begin{figure*}[!t]
 \centering
 \includegraphics[width=0.93\textwidth]{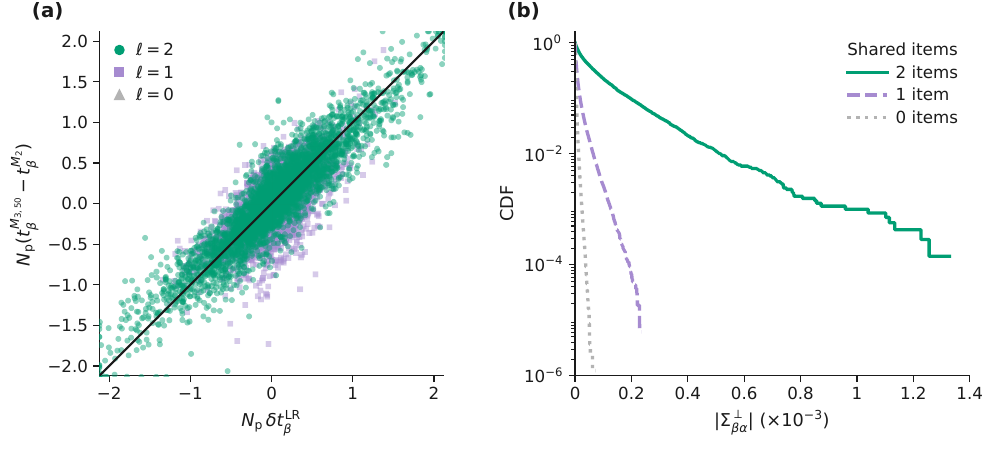}
 \caption{Effective linear response accounting for reoptimization of lower-order parameters and regularization. (a) Comparison of the predicted change in the three-item co-occurrence count from linear response, $N_{\mathrm p}\delta t_\beta^{\mathrm{LR}}$ (horizontal axis), with the change obtained by refitting the model, $N_{\mathrm p}(t_\beta^{M_{3,50}}-t_\beta^{M_2})$ (vertical axis), for the fixed set of 19,550 unselected triplets. We define $\ell(\beta)=\max_{\alpha\in S_{50}}|\alpha\cap\beta|$, and green circles, purple squares, and gray triangles indicate $\ell=2,1,$ and $0$, respectively. The black line is the identity line. Statistics are calculated from all 19,550 points; the Pearson correlation coefficient is 0.850 and the RMSE relative to the identity line is 0.186 counts. (b) CDF of $|\Sigma^\perp_{\beta\alpha}|$ over all 977,500 pairs of selected triplets $\alpha\in S_{50}$ and unselected triplets $\beta\in\Rset$, grouped by the number of shared items $|\alpha\cap\beta|=2,1,$ or $0$. Here, CDF denotes the upper-tail probability $P(|\Sigma^\perp_{\beta\alpha}|\geq s)$. The green solid, purple dashed, and gray dotted curves represent two shared items, one shared item, and no shared items, respectively. The denominator for each group includes all pairs, including those with zero coefficients.}
 \label{fig:response}
\end{figure*}

Figure~\ref{fig:response}(a) shows that the first-order changes calculated from pairwise-model covariances agree approximately with those obtained by refitting the three-body model. Across the 19,550 unselected triplets, the Pearson correlation coefficient is 0.850, the slope of the fitted regression line is 0.861, and the root-mean-square error relative to the identity line is 0.186 counts. This agreement indicates that the linear response based on the original pairwise model captures the changes in system-wide predictions induced by the added three-body couplings.

The magnitude of the effective response depends strongly on how many items the triplets share. Here the overlap classes refer to each selected-unselected pair through $|\alpha\cap\beta|$, rather than the maximum $\ell(\beta)$ over the selected set. Green, purple, and gray represent two, one, and zero shared items, respectively. Figure~\ref{fig:response}(b) compares the distributions of the absolute effective response coefficients $|\Sigma^\perp_{\beta\alpha}|$ for selected-unselected triplet pairs sharing two, one, or zero items. The plotted CDF is defined here as the upper-tail probability $P(|\Sigma^\perp_{\beta\alpha}|\geq s)$. The mean absolute coefficients, taken over all pairs in each group, are $7.33\times10^{-5}$, $7.35\times10^{-6}$, and $1.85\times10^{-6}$, respectively. The mean for pairs sharing two items is 9.97 times that for pairs sharing one item, indicating a particularly strong response between triplets sharing an item pair.

\begingroup
\color{black}
\section{Discussion and conclusions}
We have identified higher-order statistical dependencies in combinations of 50 items listed in perfume descriptions that are not reproduced by a pairwise maximum-entropy model. Even a model containing only pairwise couplings can generate correlations among three items through indirect dependencies mediated by multiple pairwise interactions. Thus, nonrandom three-item co-occurrence patterns alone do not establish the contribution of explicit three-body interactions. We instead calculated the eight-state marginal distribution of each triplet from a pairwise model inferred from the one- and two-body statistics of the full 50-item system and compared it with the observed distribution. Although the model approximately reproduces the distribution of the number of listed items per perfume, the three-item distributions exhibit substantial discrepancies between the observed and synthetic datasets. These discrepancies persist after refitting the pairwise model to each synthetic dataset and cannot be accounted for by generative controls allowing listing tendencies to vary across release-year or brand groups. They also persist after merging similar perfumes. The higher-order structure unaccounted for by the pairwise model is therefore apparent in the full three-item marginal distributions, not merely in isolated triplets.

The deviating triplets also exhibit a distinct organization. For both positive and negative deviations, triplets are concentrated around particular shared item pairs, more strongly than in datasets generated by the pairwise model. Randomizing the deviating triplets while preserving each item's participation count does not reproduce this concentration. The excess cannot therefore be explained solely by a few items appearing in many triplets. Instead, deviations repeatedly occur among triplets that share the same item pair but differ in their third item. For example, ten positively deviating triplets share Lemon and Leather. This organization is not captured by considering individual-item frequencies alone. Examining the arrangement of deviating triplets, rather than treating each triplet separately, reveals a structured pattern of higher-order dependencies.

Introducing a small number of three-body couplings also improves predictions for triplets without added couplings. When we add couplings for 50 strongly deviating triplets and reoptimize all parameters, including the one- and two-body parameters, the sum of squared residuals over the remaining 19,550 triplets decreases by 5.44\%. The improvement depends on item sharing: residual reductions are 12.53\% for triplets sharing two items with a selected triplet and 2.15\% for those sharing one item. Thus, the changes induced by the added couplings are not uniform across the system but are particularly strong around shared item pairs. Related approaches to representing dependencies beyond pairwise relations have been discussed in higher-order network research~\cite{Battiston_2020,Lambiotte_2019}.

These changes in predictions can be characterized quantitatively from covariances of the original pairwise model. Because the one- and two-body parameters are readjusted when three-body couplings are added, we use an effective static linear response incorporating this readjustment. The predicted first-order changes agree approximately with those obtained by refitting the three-body model, with a Pearson correlation coefficient of 0.850. The effective response coefficients also depend strongly on the number of shared items, with a mean absolute coefficient for triplet pairs sharing two items 9.97 times that for pairs sharing one item. This calculation uses only the pairwise-model covariances, observed residuals of the selected triplets, and regularization coefficients, without requiring the refitted three-body model as an input. Its agreement with the refitting results indicates that pairwise-model covariance structure, together with lower-order parameter readjustment, approximately accounts for how added three-body couplings alter predictions for other triplets.

The extent to which interaction terms of different orders are needed to describe the listed-item combinations in perfumes remains an open question. Our dataset contains 1,979 perfumes, whereas 50 items define 19,600 triplets and 230,300 quadruplets. As the interaction order increases, the number of observations supporting each combination becomes increasingly limited, making an exhaustive estimation of four-body and higher-order interactions difficult. Within this finite-sample setting, we have used the full 50-variable pairwise model as a reference, characterized the magnitude and organization of deviations in three-item distributions, and investigated the system-wide response to adding couplings for 10-50 triplets. The reduction in residuals for unselected triplets does not, by itself, uniquely determine the interaction order needed to describe the data. How much of the combination structure can be accounted for by including more three-body couplings or higher-order interactions remains to be investigated. Human observers have been shown to perceive third-order statistical features in visual textures \cite{Victor_2013}. An important next question is whether the higher-order structure identified in perfume item combinations is perceptible to humans and, if so, how it is reflected in their impressions of the fragrances. The approach of examining both the organization of deviations from a pairwise model and the responses to higher-order couplings could also be applied to other cultural combination data, including recipes.

In summary, the listed-item combinations in perfumes exhibit higher-order statistical dependencies that a pairwise model does not reproduce. Deviating triplets concentrate around shared item pairs, identifying item pairs as important units for understanding this higher-order organization. Introducing a small number of three-body couplings improves predictions for many triplets without added couplings, and these changes can be approximately described by a linear response incorporating pairwise-model covariances and the readjustment of lower-order parameters. By characterizing not only deviations from a pairwise model but also their organization and the response of other combinations to selected three-body couplings, this study provides a framework for investigating the statistical structure of perfume item combinations beyond pairwise descriptions.

\endgroup
\begin{acknowledgments}
We thank Kunihide Hoshino and Hiroshi Nagayasu for valuable discussions and expert advice. Y.S. acknowledges support from the Top Runners in Strategy of Transborder Advanced Researches (TRiSTAR) program conducted as the Strategic Professional Development Program for Young Researchers by the Ministry of Education, Culture, Sports, Science and Technology (MEXT).

The authors used ChatGPT (OpenAI) and Claude (Anthropic) to assist with generating and revising code and with drafting and revising the manuscript. The authors reviewed and revised the resulting materials and take responsibility for the scientific content of this work.
\end{acknowledgments}

\noindent\textbf{Author contributions.} \textbf{Yohei Shida:} Conceptualization, Methodology, Formal analysis, Validation, Writing - original draft. \textbf{Jun'ichi Ozaki:} Methodology, Writing - review \& editing, Supervision.

\noindent\textbf{Data availability.} The perfume data analyzed in this study were obtained with permission from Cahiers de Mode~\cite{CahiersDeMode_PerfumeEncyclopedia} and cannot be redistributed by the authors. Requests for access should be directed to Cahiers de Mode.

\appendix
\begingroup
\def\PREappendixcolor{black}
\color{\PREappendixcolor}
\makeatletter
\let\PREpendingfloatboxreset\@floatboxreset
\def\@floatboxreset{\PREpendingfloatboxreset\color{\PREappendixcolor}}
\makeatother
\let\PREpendingincludegraphics\includegraphics
\renewcommand{\includegraphics}[2][]{\begingroup\color{black}\PREpendingincludegraphics[#1]{#2}\endgroup}
\section{Data construction, model estimation, and residual evaluation}\label{sec:app_data}
We analyzed 1,979 perfumes obtained from perfume-description articles on Cahiers de Mode in June 2026. We combined the items listed in the top, middle, and base categories of each article into a single set and represented the presence or absence of each item by a binary variable for each product. Brand names and release years obtained from the same articles were used in the control analyses of between-group heterogeneity and duplicated similar products.

Of the original 2,036 item labels, we excluded six and consolidated synonyms and spelling variants among the remaining 2,030 labels to obtain 1,307 normalized items. Multiple occurrences of an item within the same product were counted only once. From the 50 most frequently listed items, we constructed a binary matrix $X\in\{0,1\}^{1979\times 50}$. At least one of these 50 items was listed for 1,943 perfumes; the remaining 36 perfumes were retained in the dataset as zero vectors over the 50 variables. The correspondence between original and normalized item labels is given in Table~\ref{tab:item_name_mapping} of Appendix~\ref{sec:app_item_names}.

 The pairwise-model estimation was initialized with $h_i=\log[p_i/(1-p_i)]$ and $J_{ij}=0$. Model expectations were evaluated by Gibbs sampling. We used Rao-Blackwellization~\cite{Gelfand_1990,Casella_Robert_1996} to estimate moments and the Adam algorithm~\cite{kingma2015adammethodstochasticoptimization} to update the parameters. For the three-body models, we initialized the parameters with the fitted pairwise-model values and $\kappa_\alpha=0$, and then refitted the one-, two-, and three-body parameters.

Each synthetic dataset consisted of 1,979 configurations of the 50 binary variables, matching the observed sample size. For the item-count distribution in Fig.~\ref{fig:k}, we used 200 synthetic datasets generated with the fitted pairwise-model parameters held fixed. For the triplet-deviance comparison in Fig.~\ref{fig:GA}, we refitted the pairwise model separately to each synthetic dataset and used the 194 datasets satisfying the fitting termination criterion. Model expectations were evaluated using MC samples independent of the chains used for model fitting.

For the residual evaluation on unselected triplets, we used eight independent MC evaluation samples from the pairwise model and four from each three-body model. The standardized residuals share a denominator based on the same pairwise-model reference probability for all models. We imposed a lower bound of $1/(2\times200\,000)$ on this probability estimated from a reference MC sample of 200,000 states. We squared the residual for each evaluation sample before averaging over samples and then summing over the triplet set.

For $m=50$, the sum of squared residuals decreased from 13,301.93 to 12,577.75, a reduction of 724.18.

\section{Supplementary evaluation and numerical treatment of triplet deviance}\label{sec:app_GA}
For the deviance $G_\alpha^{M_2}$ defined in Eq.~(\ref{eq:GA}), we examined the correspondence between observed and pairwise-model-predicted three-item co-occurrence counts and the dependence of the number of deviating triplets on the comparison threshold. Figure~\ref{fig:s_threshold}(a) shows the observed count $o_\alpha$ and predicted count $e_\alpha^{M_2}$ for all 19,600 triplets. Figure~\ref{fig:s_threshold}(b) shows the fraction of triplets with deviance at or above threshold $\tau$. The model curves and bands represent the mean and the 2.5th to 97.5th percentiles across the 194 synthetic datasets for which the pairwise model was refitted. The threshold-dependent counts are listed in Table~\ref{tab:threshold}.

\begin{table*}[t]
 \centering
 \caption{Threshold dependence of the numbers of triplets satisfying $G_\alpha^{M_2}\geq\tau$. The synthetic-data values are means across the 194 datasets evaluated using refitted pairwise models.}
 \label{tab:threshold}
 \small
 \begin{tabular}{rrrrr}
 \toprule
 Threshold $\tau$&\shortstack{Observed\\total}&\shortstack{Observed\\positive}&\shortstack{Observed\\negative}&\shortstack{Synthetic-data mean\\(total)}\\
 \midrule
 3.84&1,634&930&704&1,053.48\\
 6.63&464&287&177&203.86\\
 10.83&62&40&22&19.36\\
 15.14&10&5&5&1.80\\
 \bottomrule
 \end{tabular}
\end{table*}

\begin{figure*}[!t]
 \centering
 \includegraphics[width=0.81\textwidth]{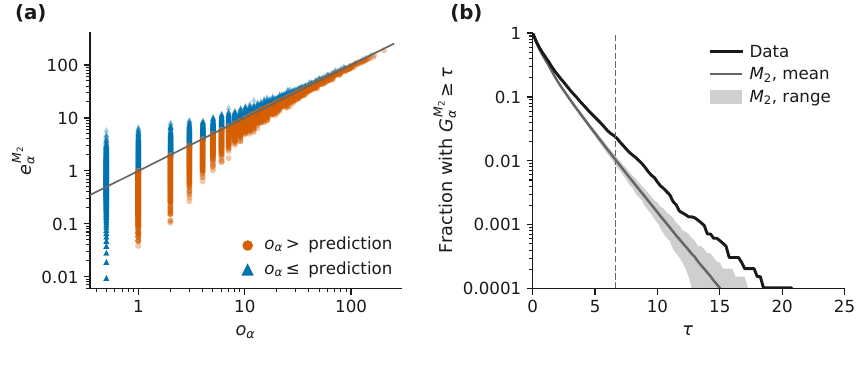}
 \caption{Three-item co-occurrence predictions of the pairwise model and the dependence of triplet deviance on the comparison threshold. (a) Observed co-occurrence counts $o_\alpha$ and pairwise-model predictions $e_\alpha^{M_2}$ for all 19,600 triplets. Orange circles and blue triangles indicate $o_\alpha>e_\alpha^{M_2}$ and $o_\alpha\leq e_\alpha^{M_2}$, respectively. The gray line is the identity line. The 3,849 triplets with zero observed co-occurrences are plotted at 0.5 on the logarithmic horizontal axis. (b) Fraction of triplets satisfying $G_\alpha^{M_2}\geq\tau$. The black curve represents the observed data; the gray curve and shaded band show the mean and 2.5th to 97.5th percentile interval, respectively, across 194 synthetic datasets evaluated after pairwise-model refitting. The gray dashed vertical line marks the reference threshold $\tau=6.63$.}
 \label{fig:s_threshold}
\end{figure*}

We estimated the three-variable marginal probabilities of the pairwise model using the MC evaluation samples described in Appendix~\ref{sec:app_data}. A pairwise model with finite parameters assigns positive marginal probabilities, but probability estimates based on finite MC samples can be zero.

Triplets for which a zero estimated probability prevented a finite numerical evaluation of $G_\alpha^{M_2}$ were assigned $G_\alpha^{M_2}=+\infty$ numerically and included in the counts above the comparison threshold. By contrast, mean deviance values were calculated over triplets with finite numerical values only. Accordingly, Fig.~\ref{fig:GA}(a) shows the mean finite deviance for each synthetic dataset, whereas the above-threshold counts in Fig.~\ref{fig:GA}(b) include triplets assigned numerical infinity. For the observed data, all 19,600 triplets had finite deviance values and were included in the mean.

\def\PREappendixcolor{black}
\color{\PREappendixcolor}
\section{Derivation of the effective linear response to three-body couplings}\label{sec:app_response}
To determine how the lower-order parameters change upon introducing three-body couplings, we linearize the regularized stationarity conditions of the three-body model around the pairwise model. Using the vectors $\bm\phi$, $\bm\theta$, $\bm\psi_S$, and $\bm\kappa_S$ defined in the main text, and denoting their empirical averages by $\widehat{\bm\phi}$ and $\widehat{\bm\psi}_S$, respectively, the stationarity conditions after reoptimization are
\begin{equation}
 \begin{aligned}
 \widehat{\bm\phi}-\avg{\bm\phi}_{\bm\theta,\bm\kappa_S}
 -\Gamma\bm\theta&=0,\\
 \widehat{\bm\psi}_S-\avg{\bm\psi_S}_{\bm\theta,\bm\kappa_S}
 -\gamma_\kappa\bm\kappa_S&=0
 \end{aligned}
 \label{eq:s_joint_stationarity}
\end{equation}
Here, $\Gamma$ applies no penalty to one-body parameters and $L_2$ regularization only to pairwise couplings.

Let $\bm\theta_{M_2}$ denote the pairwise-model parameter vector, and write $\delta\bm\theta=\bm\theta-\bm\theta_{M_2}$ and $\delta\bm\kappa_S=\bm\kappa_S$. Treating the pairwise model as a stationary point of the penalized likelihood and expanding the model expectations to first order around that point yields, with the covariance blocks $F$, $B$, and $C$ defined in the main text,
\begin{align}
 \delta\avg{\bm\phi}&\simeq F\delta\bm\theta+B\delta\bm\kappa_S,\nonumber\\
 \delta\avg{\bm\psi_S}&\simeq B^\mathsf T\delta\bm\theta+C\delta\bm\kappa_S
 \label{eq:s_moment_linear}
\end{align}
Subtracting the stationarity conditions of the pairwise model from those after reoptimization gives Eq.~(\ref{eq:s_joint_linear_system}) of the main text. Its upper block yields
\begin{equation}
 \delta\bm\theta\simeq-(F+\Gamma)^{-1}B\delta\bm\kappa_S
 \label{eq:s_dtheta}
\end{equation}
The one- and two-body parameters are therefore reoptimized when three-body couplings are introduced. For the regularized pairwise components, the difference between the stationarity conditions gives $\delta\langle\bm\phi\rangle=-\Gamma\delta\bm\theta$, so the lower-order moments are not held exactly fixed.

Substituting Eq.~(\ref{eq:s_dtheta}) into the three-body stationarity condition and using the Schur complement $\Sigma^\perp_{SS}$ defined in the main text gives
\begin{equation}
 (\Sigma^\perp_{SS}+\gamma_\kappa I_m)
 \delta\bm\kappa_S\simeq\bm r_S
 \label{eq:s_kapparesp}
\end{equation}
Here, $\bm r_S=\widehat{\bm\psi}_S-\langle\bm\psi_S\rangle_{M_2}$ is the difference between the empirical co-occurrence frequencies of the selected triplets and their expectations under the pairwise model.

The first-order change for an unselected triplet is the product of its effective response $\Sigma^\perp_{\beta S}$ and $\delta\bm\kappa_S$. Substituting Eq.~(\ref{eq:s_kapparesp}) yields Eq.~(\ref{eq:s_closed_response}) of the main text. The change in predictions for unselected triplets after refitting can thus be approximated using the pairwise-model covariances and the empirical residuals of the selected triplets. If $\Gamma=0$ and $F$ is invertible, $\Sigma^\perp_{SS}$ is the residual covariance of the three-body statistics after removing their orthogonal projection onto the lower-order statistics. In the present study, $\Gamma\neq0$; accordingly, we use an effective response that incorporates regularized reoptimization of the lower-order parameters, rather than an unregularized orthogonal projection.

We estimated the covariances from 400,000 independent MC state samples generated by the pairwise model. No samples from the three-body model were used in this covariance calculation. Taking the selected set to be $S_{50}$, we calculated the first-order response from the difference between the observed mean of the selected triplets and its pairwise-model expectation. For comparison, the changes obtained after refitting were calculated by averaging the predicted probabilities separately over four evaluation samples of the three-body model $M_{3,50}$ and eight evaluation samples of the pairwise model, and then taking the difference.

\def\PREappendixcolor{black}
\color{\PREappendixcolor}
\section{Between-group heterogeneity and merging of similar products}\label{sec:app_heterogeneity}
\subsection{Heterogeneity across release-year and brand groups}
If item-listing tendencies vary across release years or brands, a mixture of groups can exhibit three-item structure that cannot be reproduced by a single pairwise model, even when the distribution within each group follows a pairwise model. To assess this potential effect of between-group heterogeneity, we used a generative model with group-specific one-body parameters and pairwise couplings shared across all groups:
\begin{equation}
 P_{\mathrm{group}}(\bm{x}\mid g)=Z_g^{-1}\exp\!\left[
 \sum_i h_i^{(g)}x_i+\sum_{i<j}J_{ij}x_ix_j\right].
 \label{eq:s_group_gen}
\end{equation}
Here, $g$ denotes a release-year or brand group, $h_i^{(g)}$ is the one-body parameter for item $i$ in group $g$, $J_{ij}$ is a pairwise coupling common to all groups, and $Z_g$ is the group-dependent partition function. Writing $g_a$ for the group of product $a$ and $N_{\mathrm p}=1979$ for the number of products, we estimated these parameters jointly by maximizing the penalized average log-likelihood
\begin{equation}
 \begin{aligned}
 \mathcal L_{\mathrm{group}}&=\frac{1}{N_{\mathrm p}}\sum_{a=1}^{N_{\mathrm p}}
 \log P_{\mathrm{group}}(\bm{x}^{(a)}\mid g_a)\\
 &\quad-\frac{\gamma_J}{2}\sum_{i<j}J_{ij}^{2}.
 \end{aligned}
 \label{eq:s_group_penalty_v3}
\end{equation}
The one-body parameters were not regularized, and differences in group sizes were accounted for in the likelihood. We divided release years into three groups: before 2000, 2000-2014, and 2015 onward. The one perfume with an unknown release year was assigned to the middle group. For brands, we used 11 groups consisting of the ten brands with the largest numbers of listed products and all remaining brands combined. Separate group-specific generative models were fitted for the release-year and brand classifications.

From each fitted generative model, we generated synthetic datasets of 1,979 configurations of the 50 binary variables while matching the numbers of products in the observed groups. We generated 100 synthetic datasets for each of the release-year and brand conditions, and refitted an ordinary pairwise model without group information to each dataset. The regularization and fitting termination criterion were the same as in the main analysis; only datasets satisfying the termination criterion were retained for the final comparison. Among the 94 retained datasets for the release-year model and 95 for the brand model, the mean numbers of triplets with $G_\alpha^{M_2}\geq6.63$ were 204.3 and 204.9, respectively. Neither condition reproduces the observed count of 464 (Table~\ref{tab:heterogeneity}).

\subsection{Merging of similar products}
Repeated listing of similar products within the same brand could produce an apparent excess of particular item combinations. To assess this effect, we also refitted the pairwise model and evaluated triplet deviances after merging similar products.

We measured similarity using all 1,307 normalized items, rather than only the 50 selected for the main analysis. Let $\mathcal I_a$ denote the set of listed items for product $a$. For products $a$ and $b$ of the same brand, we define their Jaccard similarity as
\begin{equation}
 s_{ab}^{\mathrm{Jac}}=\frac{|\mathcal I_a\cap\mathcal I_b|}{|\mathcal I_a\cup\mathcal I_b|}
 \label{eq:s_jaccard_v3}
\end{equation}
For pairs of products with nonempty item sets, we connected products when $s_{ab}^{\mathrm{Jac}}\geq0.5$. We then retained one representative product from each connected component, choosing the product with the earliest release year as the representative.

This procedure removed 99 products, leaving a dataset of 1,880 perfumes. Refitting the pairwise model to the original 50-variable listing vectors of the retained products yielded 372 triplets with $G_\alpha^{M_2}\geq6.63$. We generated 100 synthetic datasets of 1,880 configurations each from the pairwise model fitted to the merged dataset and refitted an ordinary pairwise model to each. Among the 97 datasets satisfying the fitting termination criterion, the mean number of deviating triplets was 202.3. Thus, the triplet deviations in the observed data remain larger than those in synthetic datasets even after merging similar products (Table~\ref{tab:heterogeneity}).

\begin{table*}[t]
 \centering
 \caption{Control comparisons of the number of triplets satisfying $G_\alpha^{M_2}\geq6.63$. The synthetic-dataset means and standard deviations are calculated after refitting the pairwise model for each condition, using only the datasets retained for the final comparison.}
 \label{tab:heterogeneity}
 \small
 \begin{tabular}{lrrr}
 \toprule
 Control condition&Observed data&\shortstack{Synthetic datasets\\retained}&\shortstack{Synthetic-data mean\\$\pm$ SD}\\
 \midrule
 Main analysis (50 variables)&464&194&203.9$\,\pm\,$17.0\\
 Release-year group model&464&94&204.3$\,\pm\,$15.5\\
 Brand group model&464&95&204.9$\,\pm\,$16.9\\
 Similar products merged (1,880 perfumes)&372&97&202.3$\,\pm\,$16.7\\
 \bottomrule
 \end{tabular}
\end{table*}

\def\PREappendixcolor{black}
\color{\PREappendixcolor}
\section{Supplementary analysis of triplet deviations using local pairwise models}\label{sec:app_localG}
\subsection{Construction of the local pairwise model}
In the main text, we obtained the distribution of each triplet by marginalizing the full 50-variable pairwise model. Here, we use a different reference distribution, namely a local maximum-entropy model that preserves the empirical one- and two-body listing frequencies of each triplet, to examine deviations in the three-item distributions~\cite{Schneidman_2003}.

For a triplet $\alpha=\{i,j,k\}$, let $\bm{s}=(s_i,s_j,s_k)\in\{0,1\}^3$ denote its listing state and $n_\alpha(\bm{s})$ its empirical count. The one-body listing frequencies $p_u$ and pairwise co-occurrence frequencies $p_{uv}$ are the empirical values defined in Eq.~(\ref{eq:lowmom}). Let $\mathcal Q_\alpha$ denote the set of probability distributions $q(\bm{s})$ over the eight states satisfying nonnegativity, normalization, and the constraints $\sum_{\bm{s}}q(\bm{s})s_u=p_u$ for $u\in\alpha$ and $\sum_{\bm{s}}q(\bm{s})s_u s_v=p_{uv}$ for $u<v$ in $\alpha$. The local pairwise maximum-entropy distribution is
\begin{equation*}
 q_\alpha^{\mathrm{loc}}=\underset{q\in\mathcal Q_\alpha}{\operatorname{arg\,max}}
 \left[-\sum_{\bm s\in\{0,1\}^3}q(\bm s)\log q(\bm s)\right]
\end{equation*}
Normalization and the six constraints on one- and two-body frequencies generally leave one degree of freedom in an eight-state distribution. We determine this remaining degree of freedom by maximizing the entropy.

When the solution assigns positive probability to all eight states, it can be expressed in exponential form using only one- and two-body terms:
\begin{equation}
\begin{split}
 q_\alpha^{\mathrm{loc}}(\bm{s})={}&(Z_\alpha^{\mathrm{loc}})^{-1}
 \exp\!\Bigl[\sum_{u\in\alpha}a_u^{(\alpha)}s_u\Bigr.\\
 &\Bigl.{}+\sum_{\substack{u<v\\u,v\in\alpha}}
 b_{uv}^{(\alpha)}s_us_v\Bigr]
\end{split}
 \label{eq:s_local_model}
\end{equation}
Here, $a_u^{(\alpha)}$ and $b_{uv}^{(\alpha)}$ are the one- and two-body parameters of the local model, respectively, and $Z_\alpha^{\mathrm{loc}}$ is its eight-state partition function. When the empirical lower-order frequencies restrict some admissible states to zero probability, the local model remains defined by constrained entropy maximization with zero probabilities permitted.

With $N_{\mathrm p}=1979$, we define the deviance relative to the local model as
\begin{equation}
 G_\alpha=2\sum_{\bm{s}\in\{0,1\}^3}n_\alpha(\bm{s})
 \log\frac{n_\alpha(\bm{s})}{N_{\mathrm p}q_\alpha^{\mathrm{loc}}(\bm{s})}
 \label{eq:s_localG}
\end{equation}
Terms with $n_\alpha(\bm{s})=0$ are taken to be zero. Using the observed three-item co-occurrence count $o_\alpha=n_\alpha(1,1,1)$ and its local-model prediction $e_\alpha=N_{\mathrm p}q_\alpha^{\mathrm{loc}}(1,1,1)$, we define positive deviations by $o_\alpha>e_\alpha$ and negative deviations by $o_\alpha<e_\alpha$.

\subsection{Deviations from local pairwise models}
For the observed data and 200 synthetic datasets generated from the full 50-variable pairwise model, each containing 1,979 configurations, we constructed a local model and calculated its deviance $G_\alpha$ separately for all 19,600 triplets. For each synthetic dataset, the local model was constructed by fixing the empirical one- and two-body listing frequencies of that synthetic dataset. \textbf{The full 50-variable pairwise model was not refitted separately to the synthetic datasets in this supplementary analysis.}

Deviations in the three-item distributions remain larger in the observed data than in the synthetic datasets when local pairwise models are used as the reference. In Fig.~\ref{fig:s_localG}(a), the mean deviance is 1.272 for the observed data and 1.059 on average across the 200 synthetic datasets. Figure~\ref{fig:s_localG}(b) shows that 279 positively and 131 negatively deviating triplets in the observed data exceed the reference threshold $G_\alpha\geq6.63$, compared with respective means of 142.0 and 73.1 in the synthetic datasets. As in the main analysis using the full 50-variable model, deviations persist in both directions under the local reference. However, because the reference distributions differ, the predicted counts and deviances of individual triplets are distinct from those in the main analysis.

Under the local pairwise reference, observed three-item co-occurrence counts also display deviations in both directions. Figure~\ref{fig:s_localG}(c) shows triplets lying outside the 95\% and 99.8\% empirical bands obtained from synthetic datasets. For Vetiver, Ylang-ylang, and Aldehydes in Fig.~\ref{fig:s_localG}(d), the observed count is 24, compared with a local-model prediction of 14.85, giving $G_\alpha=21.15$. For Leather, Pepper, and Saffron, the observed count is zero despite a local-model prediction of 6.34, giving $G_\alpha=21.51$.

\begin{figure*}[!t]
 \centering
 \includegraphics[width=0.93\textwidth]{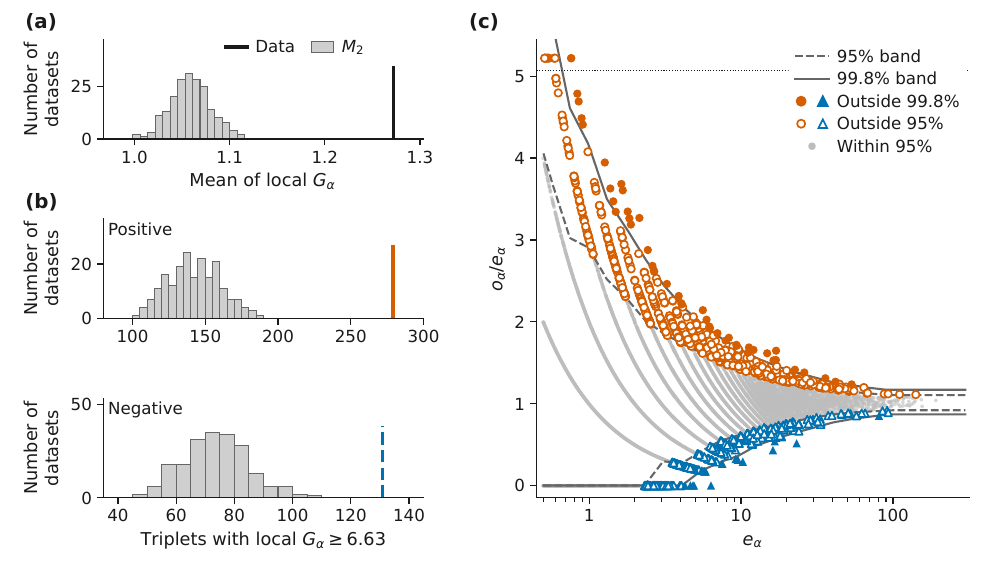}\\[1mm]
 \includegraphics[width=0.72\textwidth]{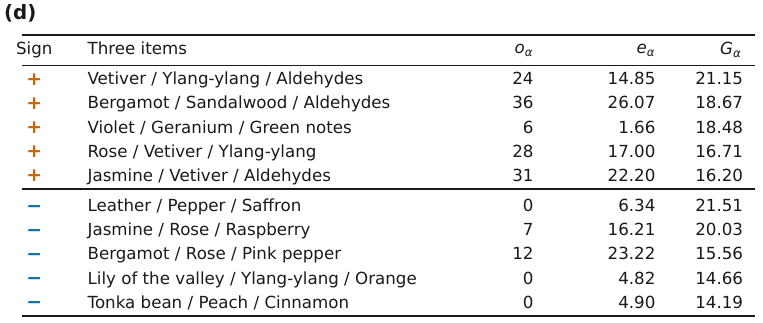}
 \caption{Deviations in three-item distributions relative to local pairwise models. We use the local-model deviance $G_\alpha$ obtained by preserving the empirical one- and two-body statistics of each triplet and present comparisons corresponding to those in Fig.~\ref{fig:GA}. (a) Mean local deviance $G_\alpha$ over all 19,600 triplets. (b) Numbers of positive and negative triplets with $G_\alpha\geq6.63$. The gray histograms in (a) and (b) are obtained from 200 synthetic datasets generated by the full pairwise model \textbf{without refitting that 50-variable model}. Vertical lines indicate the observed data; in (b), the orange solid and blue dashed lines indicate positive and negative deviations, respectively. (c) Funnel plot of the observed-to-predicted ratio $o_\alpha/e_\alpha$ for the 16,428 triplets with local-model predicted counts $e_\alpha\geq0.5$. Dashed and solid curves indicate the 95\% and 99.8\% empirical bands obtained from synthetic datasets, respectively. Orange circles and blue triangles represent positive and negative deviations; filled symbols lie outside the 99.8\% band, open symbols lie outside the 95\% band but within the 99.8\% band, and light gray points lie within the 95\% band. The nine triplets with ratios exceeding 5 are displayed above the upper dotted line. (d) Five triplets with large local deviance for each deviation sign. Here, $o_\alpha$ is the observed three-item co-occurrence count and $e_\alpha$ is the count predicted by the local model.}
 \label{fig:s_localG}
\end{figure*}

\clearpage
\onecolumngrid
\section{Correspondence between original and normalized item labels}\label{sec:app_item_names}
Of the 2,036 original item labels, six were excluded, and the remaining 2,030 labels were consolidated into 1,307 normalized items. Table~\ref{tab:item_name_mapping} lists 317 items formed by combining multiple original labels and 276 items renamed from a single original label. These 593 normalized items correspond to 1,316 original labels. The remaining 714 items, each represented by a single unchanged original label, are not included in the table. The original labels are reproduced in Japanese as recorded in the source data.
\begingroup
\footnotesize
\setlength{\tabcolsep}{4pt}
\renewcommand{\arraystretch}{1.12}
\setlength{\LTcapwidth}{\textwidth}
\begin{longtable}{>{\raggedright\arraybackslash}p{0.36\textwidth}>{\raggedright\arraybackslash}p{0.58\textwidth}}
\caption{Correspondence between normalized item names and original labels. The table lists 317 items formed by combining multiple original labels and 276 items renamed from a single original label. Original labels are reproduced as recorded in the Japanese source data.}\label{tab:item_name_mapping}\\
\toprule
Normalized item name & Original item labels (Japanese source) \\
\midrule
\endfirsthead
\caption[]{Correspondence between normalized item names and original labels (continued).}\\
\toprule
Normalized item name & Original item labels (Japanese source) \\
\midrule
\endhead
\midrule
\multicolumn{2}{r}{Continued on next page}\\
\endfoot
\bottomrule
\endlastfoot
Jasmine & \mbox{イタリア＆モロッコ・ジャスミン};\allowbreak\ \mbox{ジャスミン};\allowbreak\ \mbox{ジャスミン・サンバック};\allowbreak\ \mbox{エジプト産ジャスミン};\allowbreak\ \mbox{ジャスミンサンバック};\allowbreak\ \mbox{グラース産ジャスミン};\allowbreak\ \mbox{インド産ジャスミン};\allowbreak\ \mbox{インド産ジャスミン・サンバック};\allowbreak\ \mbox{モロッコ産ジャスミン};\allowbreak\ \mbox{インディアン・ジャスミン};\allowbreak\ \mbox{インド産ジャスミンサンバック};\allowbreak\ \mbox{中国産ジャスミン・サンバック};\allowbreak\ \mbox{イタリアン・ジャスミン};\allowbreak\ \mbox{イタリア産ジャスミン};\allowbreak\ \mbox{エジプト産ジャスミン・サンバック};\allowbreak\ \mbox{中国産ジャスミンサンバック};\allowbreak\ \mbox{イタリア産ジャスミン・グランディフローラム};\allowbreak\ \mbox{インド産ジャスミン・グランディフローラム};\allowbreak\ \mbox{グラース産ジャスミン・グランディフローラム};\allowbreak\ \mbox{マダガスカル産ジャスミン} \\
SFE incense extract & \mbox{SFE インセンス・エクストラクト} \\
SFE ginger extract & \mbox{SFE ジンジャー・エキス} \\
Gardenia & \mbox{くちなし（ガーデニア）};\allowbreak\ \mbox{タヒチアン・ガーデニア};\allowbreak\ \mbox{カリフォルニア産ガーデニア};\allowbreak\ \mbox{ガーデニア};\allowbreak\ \mbox{ガーデニア（クチナシ）};\allowbreak\ \mbox{タヒチ産ガーデニア};\allowbreak\ \mbox{ブラジル産ガーデニア};\allowbreak\ \mbox{クチナシ} \\
Sugar cane & \mbox{さとうきび};\allowbreak\ \mbox{サトウキビ} \\
Lily of the valley & \mbox{すずらん};\allowbreak\ \mbox{スズラン};\allowbreak\ \mbox{フランス産スズラン};\allowbreak\ \mbox{ミュゲ（スズラン）};\allowbreak\ \mbox{リリー・オブ・ザ・ヴァレー} \\
Orris concrete & \mbox{中国産とモロッコ産アイリス・コンクリート};\allowbreak\ \mbox{アイリス・コンクリート};\allowbreak\ \mbox{イリスコンクリート};\allowbreak\ \mbox{アイリスコンクリート};\allowbreak\ \mbox{イリス・コンクリート（イリスバター）};\allowbreak\ \mbox{オリス・コンクリート};\allowbreak\ \mbox{オリスコンクリート} \\
Hedione & \mbox{へディオン};\allowbreak\ \mbox{ヘディオン};\allowbreak\ \mbox{ヘディオン®} \\
Vetiver & \mbox{べチバー};\allowbreak\ \mbox{タヒチアン・ベチバー};\allowbreak\ \mbox{ハイチアン・ベチバー};\allowbreak\ \mbox{ベチバー};\allowbreak\ \mbox{ハイチ産ベチバー};\allowbreak\ \mbox{タヒチ産ベチバー};\allowbreak\ \mbox{ジャワ産ベチバー};\allowbreak\ \mbox{ハイチベチバー};\allowbreak\ \mbox{バーボン・ベチバー};\allowbreak\ \mbox{ブラジル産ベチバー};\allowbreak\ \mbox{ブルボン産ベチバー};\allowbreak\ \mbox{ベティバー};\allowbreak\ \mbox{インドベティバー};\allowbreak\ \mbox{インド産ベティバー};\allowbreak\ \mbox{ペチバー};\allowbreak\ \mbox{ジャワ島産ベチバー} \\
Bergamot & \mbox{べルガモット};\allowbreak\ \mbox{ベルガモット};\allowbreak\ \mbox{カラブリアン・ベルガモット};\allowbreak\ \mbox{カラブリア産ベルガモット};\allowbreak\ \mbox{イタリアン・ベルガモット};\allowbreak\ \mbox{カラブリアンベルガモット};\allowbreak\ \mbox{シチリア産ベルガモット};\allowbreak\ \mbox{シシリアン・ベルガモット};\allowbreak\ \mbox{イタリア産ベルガモット};\allowbreak\ \mbox{カリビアン・ベルガモット};\allowbreak\ \mbox{シチリアンベルガモット} \\
Ice accord & \mbox{アイス・アコード} \\
Iris & \mbox{アイリス};\allowbreak\ \mbox{フィレンツェ産アイリス};\allowbreak\ \mbox{ニオイイリス};\allowbreak\ \mbox{イリス};\allowbreak\ \mbox{トスカーナ産アイリス};\allowbreak\ \mbox{イリス（イタリア産} \\
Orris root & \mbox{ニオイイリスの根茎};\allowbreak\ \mbox{ニオイ・イリスの根茎};\allowbreak\ \mbox{イリスルート};\allowbreak\ \mbox{オリスルート};\allowbreak\ \mbox{イタリア産オリスルート};\allowbreak\ \mbox{オリス・ルート} \\
Iris absolute & \mbox{イリスアブソリュート};\allowbreak\ \mbox{イリス・アブソリュート};\allowbreak\ \mbox{トスカーナ産アイリス・アブソリュート};\allowbreak\ \mbox{アイリス・アブソリュート} \\
Orris butter & \mbox{イリスバター};\allowbreak\ \mbox{モロッコ産イリスバター};\allowbreak\ \mbox{中国産イリスバター};\allowbreak\ \mbox{オリス・バター};\allowbreak\ \mbox{オリスバター} \\
Iris pallida & \mbox{アイリスパリダ};\allowbreak\ \mbox{グラース産アイリス・パリダ};\allowbreak\ \mbox{イリスパリダ} \\
Iris flower & \mbox{アイリス・フラワー};\allowbreak\ \mbox{アイリスフラワー} \\
Iris oil & \mbox{イリス油} \\
Acacia & \mbox{アカシア};\allowbreak\ \mbox{エジプト産アカシア} \\
Acacia farnesiana & \mbox{アカシア・キンゴウカン} \\
Acacia honey & \mbox{アカシア・ハニー} \\
Agarwood & \mbox{アガーウッド（ウード）};\allowbreak\ \mbox{アガーウッド};\allowbreak\ \mbox{ラオス産アガーウッド（ウード）};\allowbreak\ \mbox{ラオス産アガーウッド};\allowbreak\ \mbox{アガーウッド（沈香）} \\
Akigalawood™ & \mbox{アキガラウッド™} \\
Aquatic accord & \mbox{アクアティック・アコード} \\
Assam oud & \mbox{アッサム・ウード} \\
Upcycled rose water & \mbox{トルコ産アップサイクルローズウォーター} \\
Apple & \mbox{アップル};\allowbreak\ \mbox{アップル（カルヴァドスアコード）} \\
Apple blossom & \mbox{アップル・ブロッサム};\allowbreak\ \mbox{アップルブロッサム} \\
Animalic notes & \mbox{アニマル・ノーツ};\allowbreak\ \mbox{アニマルノート} \\
Apricot blossom & \mbox{アプリコット・ブロッサム} \\
Amalfi lemon & \mbox{アマルフィ・レモン};\allowbreak\ \mbox{アマルフィレモン} \\
Lemon & \mbox{アマルフィ産レモン};\allowbreak\ \mbox{アルゼンチン産レモン};\allowbreak\ \mbox{シシリア産レモン};\allowbreak\ \mbox{レモン};\allowbreak\ \mbox{シチリア産レモン};\allowbreak\ \mbox{シチリアン・レモン};\allowbreak\ \mbox{イタリアン・レモン};\allowbreak\ \mbox{カラブリア産レモン};\allowbreak\ \mbox{シチリアンレモン};\allowbreak\ \mbox{イタリアンレモン};\allowbreak\ \mbox{レモン（シトロン）} \\
Alaska cedar essence & \mbox{アラスカ・シダー・エッセンス} \\
Arabian jasmine & \mbox{アラビアン・ジャスミン} \\
Arabian jasmine sambac & \mbox{アラビアン・ジャスミン・サンバック} \\
Arancia vaniglia & \mbox{アランチャ ヴァニリア} \\
Honey & \mbox{アリカンテ産ハニー};\allowbreak\ \mbox{ハニー};\allowbreak\ \mbox{ラオス産ハニー};\allowbreak\ \mbox{ハニー（蜂蜜）};\allowbreak\ \mbox{プロヴァンス産ハニー};\allowbreak\ \mbox{蜂蜜} \\
Aldehydes & \mbox{アルデハイド};\allowbreak\ \mbox{アルデヒド} \\
Armoise & \mbox{アルモワーズ（ヨモギ）} \\
Angelica seed & \mbox{アンジェリカシード};\allowbreak\ \mbox{アンジェリカ・シード} \\
Angelica root essence & \mbox{アンジェリカ・ルート・エッセンス} \\
Vanilla & \mbox{タヒチアン・バニラ};\allowbreak\ \mbox{バニラ};\allowbreak\ \mbox{マダガスカル産バニラ};\allowbreak\ \mbox{タヒチ産バニラ};\allowbreak\ \mbox{ブルボンバニラ};\allowbreak\ \mbox{バーボンバニラ};\allowbreak\ \mbox{バーボン・バニラ};\allowbreak\ \mbox{マダガスカル・バニラ};\allowbreak\ \mbox{インド産バニラ};\allowbreak\ \mbox{マダガスカルバニラ};\allowbreak\ \mbox{ブラジル産バニラ};\allowbreak\ \mbox{ブルボン・バニラ};\allowbreak\ \mbox{ベニラ};\allowbreak\ \mbox{マヨット島産バニラ};\allowbreak\ \mbox{マルキーズ諸島産バニラ} \\
Vanilla infusion & \mbox{タヒチアンバニラインフュージョン} \\
Vanilla flower & \mbox{タヒチアン・バニラ・フラワー} \\
Ambergris & \mbox{アンバーグリス};\allowbreak\ \mbox{アンバーグリス(龍涎香)};\allowbreak\ \mbox{アンバーグリス（Cachalox®）};\allowbreak\ \mbox{アンバーグリス（Cachalox）};\allowbreak\ \mbox{龍涎香} \\
Ambrette & \mbox{アンブレット};\allowbreak\ \mbox{アンブレット（ムスクシード）};\allowbreak\ \mbox{インドネシア産アンブレット} \\
Ambrette seed & \mbox{アンブレットシード};\allowbreak\ \mbox{ペルー産アンブレットシード} \\
Ambrettolide & \mbox{アンブレットリド};\allowbreak\ \mbox{アンブレットリド（ムスク）} \\
Ambroxan & \mbox{アンブロキサン};\allowbreak\ \mbox{アンブロキサン®};\allowbreak\ \mbox{アンブロクサン} \\
Ambroxide & \mbox{アンブロキシド};\allowbreak\ \mbox{アンブロキド} \\
Ambrox & \mbox{アンブロックス};\allowbreak\ \mbox{アンブロックス®} \\
Ambrofix & \mbox{アンブロフィックス};\allowbreak\ \mbox{アンブロフィックス®} \\
Ambrofix™ & \mbox{アンブロフィックス™} \\
Almond & \mbox{アーモンド};\allowbreak\ \mbox{インド産アーモンド} \\
Almond tree & \mbox{アーモンド・ツリー};\allowbreak\ \mbox{アーモンドツリー} \\
Almond blossom & \mbox{アーモンド・ブロッサム};\allowbreak\ \mbox{アーモンドブロッサム} \\
Almond milk & \mbox{アーモンドミルク};\allowbreak\ \mbox{アーモンド・ミルク} \\
Frankincense & \mbox{イエメン産フランキンセンス};\allowbreak\ \mbox{オマーン産フランキンセンス};\allowbreak\ \mbox{フランキンセンス};\allowbreak\ \mbox{ソマリア産フランキンセンス} \\
Yellow plum & \mbox{イエロー・プラム} \\
Yellow mandarin & \mbox{イエロー・マンダリン};\allowbreak\ \mbox{イエローマンダリン} \\
Cool mint & \mbox{イギリス産クールミント};\allowbreak\ \mbox{クールミント} \\
Iso E Super & \mbox{イシEスーパー};\allowbreak\ \mbox{イソEスーパー};\allowbreak\ \mbox{イソＥスーパー} \\
Orange & \mbox{イスキア島産オレンジ};\allowbreak\ \mbox{オレンジ};\allowbreak\ \mbox{シチリア産オレンジ};\allowbreak\ \mbox{イタリアン・オレンジ};\allowbreak\ \mbox{シシリアン・オレンジ};\allowbreak\ \mbox{カリフォルニア・オレンジ};\allowbreak\ \mbox{カリフォルニア産オレンジ};\allowbreak\ \mbox{フロリダ産オレンジ};\allowbreak\ \mbox{ポルトガル産オレンジ} \\
Cypress & \mbox{イトスギ};\allowbreak\ \mbox{モロッコ産イトスギ} \\
Immortelle & \mbox{イモーテル};\allowbreak\ \mbox{イモーテル（カレープラント）};\allowbreak\ \mbox{カレープラント} \\
Ylang-ylang & \mbox{イランイラン};\allowbreak\ \mbox{コモロ産イランイラン};\allowbreak\ \mbox{マダガスカル産イランイラン};\allowbreak\ \mbox{イラン・イラン};\allowbreak\ \mbox{イランライン};\allowbreak\ \mbox{マイヨット産イランイラン} \\
Ylang-ylang absolute & \mbox{イランイラン アブソリュート} \\
Ylang-ylang essence & \mbox{コモロ産イランイラン・エッセンス} \\
Ylang-ylang heart & \mbox{マダガスカル産イランイランハート} \\
English ivy & \mbox{イングリッシュ・アイビー} \\
English garden rose & \mbox{イングリッシュガーデン・ローズ} \\
English lavender & \mbox{イングリッシュ・ラベンダー} \\
Incense & \mbox{インセンス};\allowbreak\ \mbox{インドネシア産インセンス};\allowbreak\ \mbox{ソマリア産インセンス} \\
Incense Vulcain & \mbox{インセンス・バルカン} \\
Wisteria & \mbox{ウィステリア};\allowbreak\ \mbox{ウィステリア（藤）};\allowbreak\ \mbox{藤};\allowbreak\ \mbox{藤（ウィステリア）} \\
Weeping silver lime headspace & \mbox{ウィーピングシルバーライム ヘッドスペース} \\
Fennel & \mbox{ウイキョウ};\allowbreak\ \mbox{フェンネル};\allowbreak\ \mbox{フェンネル（ういきょう）} \\
Watery notes & \mbox{ウォータリー・ノート};\allowbreak\ \mbox{ウォータリーノート} \\
Water jasmine & \mbox{ウォーター・ジャスミン};\allowbreak\ \mbox{ウォータージャスミン} \\
Water notes & \mbox{ウォーター・ノーツ};\allowbreak\ \mbox{ウォーターノート};\allowbreak\ \mbox{ウォーター・ノート} \\
Water hyacinth & \mbox{ウォーターヒヤシンス};\allowbreak\ \mbox{ウォーター・ヒヤシンス} \\
Water lily & \mbox{ウォーターリリー};\allowbreak\ \mbox{ウォーター・リリー} \\
Woody notes & \mbox{ウッディ・ノーツ};\allowbreak\ \mbox{ウッディノーツ};\allowbreak\ \mbox{ウッディノート};\allowbreak\ \mbox{ウッディ・ノート};\allowbreak\ \mbox{ウッディー・ノーツ};\allowbreak\ \mbox{ウッディーノート};\allowbreak\ \mbox{ウッディー・ノート} \\
Wood & \mbox{インディアンウッド};\allowbreak\ \mbox{ウッド} \\
Ultravanil™ & \mbox{ウルトラバニル™} \\
Oud & \mbox{ウード};\allowbreak\ \mbox{ウード（アガーウッド）};\allowbreak\ \mbox{ラオス産ウード} \\
Oud accord & \mbox{ウード・アコード} \\
Oud Assam & \mbox{ウード・アッサム} \\
Oud absolute & \mbox{ウード・アブソリュート} \\
Oud wood & \mbox{ウードウッド};\allowbreak\ \mbox{ウード・ウッド} \\
Oolong tea & \mbox{ウーロン・ティー};\allowbreak\ \mbox{ウーロン茶} \\
Exaltone & \mbox{エキサルトン®} \\
Ambrette absolute & \mbox{エクアドル産とペルー産のアンブレット・アブソリュート} \\
Egyptian jasmine & \mbox{エジプシャン・ジャスミン} \\
Egyptian jasmine absolute & \mbox{エジプシャン・ジャスミン・アブソリュート} \\
Geranium & \mbox{エジプシャンゼラニウム};\allowbreak\ \mbox{ゼラニウム};\allowbreak\ \mbox{エジプト産ゼラニウム};\allowbreak\ \mbox{ブルボン産ゼラニウム};\allowbreak\ \mbox{バーボン・ゼラニウム};\allowbreak\ \mbox{アフリカン・ゼラニウム};\allowbreak\ \mbox{ブルボン・ゼラニウム} \\
Evernyl & \mbox{エベルニル（モス）} \\
Ebony & \mbox{エボニー};\allowbreak\ \mbox{エボニー（黒檀）};\allowbreak\ \mbox{コクタン（エボニー）} \\
Ebony wood & \mbox{エボニーウッド};\allowbreak\ \mbox{エボニー・ウッド} \\
Elemi & \mbox{エレミ};\allowbreak\ \mbox{マレーシア産エレミ} \\
Elemi resin & \mbox{エレミ・レジン} \\
Angelwing jasmine & \mbox{エンジェルウィング・ジャスミン} \\
Osmanthus & \mbox{オスマンサス};\allowbreak\ \mbox{中国産オスマンサス};\allowbreak\ \mbox{オスマンサス（キンモクセイ）};\allowbreak\ \mbox{オスマンサス（金木犀）} \\
Osman rose & \mbox{オスマン・ローズ} \\
Ozonic notes & \mbox{オゾンノート};\allowbreak\ \mbox{オゾン・ノート} \\
Opuntia & \mbox{オプンティア};\allowbreak\ \mbox{オプンティア（ウチワサボテン）} \\
Opoponax & \mbox{オポポナックス};\allowbreak\ \mbox{ソマリア産オポポナックス};\allowbreak\ \mbox{オポポナックス（ミルラ）} \\
Orris accord & \mbox{オリス・アコード} \\
Olibanum & \mbox{オリバナム};\allowbreak\ \mbox{オリバナム（フランキンセンス）};\allowbreak\ \mbox{ソマリア産オリバナム} \\
Olibanum Vulcain & \mbox{オリバナム・ヴァルカン} \\
Olive tree & \mbox{オリーブツリー};\allowbreak\ \mbox{オリーブ・ツリー} \\
Orcanox & \mbox{オルカノックス（アンブロキシド）} \\
Orange peel & \mbox{オレンジ・ピール};\allowbreak\ \mbox{オレンジピール} \\
Orange blossom & \mbox{アフリカン・オレンジ・フラワー};\allowbreak\ \mbox{オレンジフラワー};\allowbreak\ \mbox{アフリカン・オレンジフラワー};\allowbreak\ \mbox{アフリカンオレンジフラワー};\allowbreak\ \mbox{チュニジア産オレンジ・フラワー};\allowbreak\ \mbox{アフリカ産オレンジフラワー};\allowbreak\ \mbox{アフリカ産オレンジ・フラワー};\allowbreak\ \mbox{オレンジ・フラワー};\allowbreak\ \mbox{モロッコ産オレンジフラワー};\allowbreak\ \mbox{オレンジ・ブロッサム};\allowbreak\ \mbox{オレンジブロッサム};\allowbreak\ \mbox{チュニジア産オレンジ・ブロッサム};\allowbreak\ \mbox{チュニジア産オレンジブロッサム};\allowbreak\ \mbox{アフリカン・オレンジ・ブロッサム};\allowbreak\ \mbox{アマルフィ産オレンジブロッサム};\allowbreak\ \mbox{オレンジ・ブロッサムフラワー};\allowbreak\ \mbox{カラブリア産オレンジ・ブロッサム};\allowbreak\ \mbox{スペイン産オレンジ・ブロッサム};\allowbreak\ \mbox{チュニジアン・オレンジ・ブロッサム};\allowbreak\ \mbox{チュニジア産オレンジブロッサム(精油};\allowbreak\ \mbox{フランス産オレンジ・ブロッサム} \\
Orange blossom absolute & \mbox{オレンジフラワー・アブソリュート};\allowbreak\ \mbox{オレンジブロッサム・アブソリュート};\allowbreak\ \mbox{オレンジ・ブロッサム・アブソリュート};\allowbreak\ \mbox{チュニジア産オレンジ・ブロッサム・アブソリュート};\allowbreak\ \mbox{エジプト産オレンジブロッサム・アブソリュート} \\
Orange blossom petal extract & \mbox{オレンジ・ブロッサム・ペタル・エクストラクト} \\
Orange leaf & \mbox{オレンジ・リーフ};\allowbreak\ \mbox{アフリカン・オレンジ・リーフ};\allowbreak\ \mbox{オレンジリーフ} \\
Organic orange essential oil & \mbox{オーガニック・オレンジ精油} \\
Oakmoss & \mbox{オークモス};\allowbreak\ \mbox{マケドニア産オークモス} \\
Sandalwood & \mbox{オーストラリア産サンダルウッド};\allowbreak\ \mbox{サンダルウッド};\allowbreak\ \mbox{スリランカ産サンダルウッド};\allowbreak\ \mbox{オーストラリアン・サンダルウッド};\allowbreak\ \mbox{マイソール産サンダルウッド};\allowbreak\ \mbox{マイソール・サンダルウッド};\allowbreak\ \mbox{インド産サンダルウッド};\allowbreak\ \mbox{マイソール産のサンダルウッド（白檀）};\allowbreak\ \mbox{ニューカレドニア産サンダルウッド} \\
Austrian sandalwood ORPUR™ & \mbox{オーストリアン・サンダルウッドORPUR™} \\
Cayenne pepper & \mbox{カイエンペッパー};\allowbreak\ \mbox{カイエンペッパー（赤唐辛子）} \\
Cacao & \mbox{カカオ};\allowbreak\ \mbox{コートジボアール産カカオ} \\
Cacao absolute & \mbox{カカオアブソリュート};\allowbreak\ \mbox{カカオ・アブソリュート} \\
Cactus juice & \mbox{カクタス・ジュース} \\
Cactus flower & \mbox{カクタス・フラワー} \\
Casablanca lily & \mbox{カサブランカリリー};\allowbreak\ \mbox{カサブランカ・リリー} \\
Stone pine & \mbox{イタリアカサマツ} \\
Blackcurrant & \mbox{カシス};\allowbreak\ \mbox{カシス（ブラック・カラント）};\allowbreak\ \mbox{ブラックカラント};\allowbreak\ \mbox{ブラック・カラント};\allowbreak\ \mbox{ブラックカラント（クロスグリ）};\allowbreak\ \mbox{ブルゴーニュ産ブラックカラント} \\
Cashmere wood & \mbox{カシミアウッド};\allowbreak\ \mbox{カシミア・ウッド} \\
Cashmere musk & \mbox{カシミアムスク};\allowbreak\ \mbox{カシミア・ムスク};\allowbreak\ \mbox{カシミヤムスク} \\
Cashmeran wood & \mbox{カシュメラン・ウッド} \\
Cassie flower & \mbox{カッシー・フラワー} \\
Canada balsam & \mbox{カナダ・バルサム} \\
Cedar leaf essence & \mbox{カナダ産シダーリーフ・エッセンス} \\
Kahili ginger & \mbox{カヒリ・ジンジャー} \\
Camellia & \mbox{カメリア};\allowbreak\ \mbox{カメリア（椿）} \\
California lemon & \mbox{カリフォルニア・レモン} \\
Cardamom & \mbox{カルダモン};\allowbreak\ \mbox{インド産カルダモン};\allowbreak\ \mbox{グアテマラ産カルダモン};\allowbreak\ \mbox{セイロン産カルダモン} \\
Cardamom essence & \mbox{カルダモン・エッセンス} \\
Camphor wood & \mbox{カンファーウッド（クスノキ）} \\
Carnation & \mbox{カーネーション};\allowbreak\ \mbox{インドネシア産カーネーション};\allowbreak\ \mbox{インド産カーネーション} \\
Carla lavender & \mbox{カーラ・ラベンダー};\allowbreak\ \mbox{プロヴァンス産カーラ・ラベンダー} \\
Guaiacwood & \mbox{ガイアックウッド};\allowbreak\ \mbox{ガイアック・ウッド};\allowbreak\ \mbox{ガイヤックウッド} \\
Galbanum essence & \mbox{ガルバナム・エッセンス} \\
Gurjun balsam & \mbox{ガージャン・バルサム} \\
Kyara incense & \mbox{キャラ・インセンス（オリバナム）} \\
King William pear & \mbox{キングウィリアム・ペアー} \\
Clary sage absolute & \mbox{クラリセージ・アブソリュート};\allowbreak\ \mbox{フランス産クラリセージ・アブソリュート} \\
Clearwood & \mbox{クリアウッド};\allowbreak\ \mbox{クリアウッド®} \\
Crystal moss & \mbox{クリスタルモス};\allowbreak\ \mbox{クリスタル・モス} \\
Clementine & \mbox{クレメンタイン};\allowbreak\ \mbox{クレメンタイン（小型のオレンジ）} \\
Clementine flower & \mbox{クレメンタイン・フラワー} \\
Creme Chantilly & \mbox{クレーム・シャンティ} \\
Clove & \mbox{クローブ};\allowbreak\ \mbox{マダガスカル・クローブ};\allowbreak\ \mbox{マダガスカル産クローブ} \\
Grass & \mbox{グラス};\allowbreak\ \mbox{グラス（草）};\allowbreak\ \mbox{草} \\
Granny Smith apple & \mbox{グラニースミスアップル};\allowbreak\ \mbox{グラニースミス・アップル};\allowbreak\ \mbox{グラニースミス・アップル（アップルパイ）} \\
Green accord & \mbox{グリーンアコード};\allowbreak\ \mbox{グリーン・アコード} \\
Green apple & \mbox{グリーンアップル};\allowbreak\ \mbox{エジプト産グリーンアップル};\allowbreak\ \mbox{グリーン・アップル} \\
Green almond & \mbox{グリーンアーモンド};\allowbreak\ \mbox{グリーン・アーモンド} \\
Green tea & \mbox{グリーンティ};\allowbreak\ \mbox{グリーン・ティ};\allowbreak\ \mbox{グリーンティー};\allowbreak\ \mbox{グリーン・ティー} \\
Green nectar & \mbox{グリーン・ネクター} \\
Green notes & \mbox{グリーン・ノーツ};\allowbreak\ \mbox{グリーンノート};\allowbreak\ \mbox{グリーン・ノート} \\
Green pear & \mbox{グリーンペアー};\allowbreak\ \mbox{グリーン・ペアー} \\
Green mandarin & \mbox{グリーン・マンダリン};\allowbreak\ \mbox{グリーンマンダリン};\allowbreak\ \mbox{シチリア産グリーンマンダリン} \\
Green leaf & \mbox{グリーンリーフ};\allowbreak\ \mbox{グリーン・リーフ} \\
Green leaves & \mbox{グリーンリーブス};\allowbreak\ \mbox{グリーン・リーブス};\allowbreak\ \mbox{グリーンリーヴス};\allowbreak\ \mbox{グリーン・リーヴス} \\
Green violet & \mbox{グリーン・ヴァイオレット} \\
Grapefruit & \mbox{グレープフルーツ};\allowbreak\ \mbox{シチリア産グレープフルーツ};\allowbreak\ \mbox{シチリアン・グレープフルーツ};\allowbreak\ \mbox{スペイン産グレープフルーツ};\allowbreak\ \mbox{フロリダ産グレープフルーツ} \\
Grapefruit essence & \mbox{グレープフルーツ・エッセンス};\allowbreak\ \mbox{メキシコ産＆アメリカ産グレープフルーツ・エッセンス} \\
Grapefruit flower & \mbox{イタリアングレープフルーツフラワー} \\
Grapefruit blossom & \mbox{グレープフルーツ・ブロッサム} \\
Bluegrass & \mbox{ケンタッキー産ブルーグラス} \\
Gentian & \mbox{ゲンチアナ（リンドウ）} \\
Cocoa & \mbox{ココア};\allowbreak\ \mbox{ココア（カカオ）} \\
Coconut water & \mbox{ココナッツウォーター};\allowbreak\ \mbox{ココナッツ・ウォーター} \\
Coconut oil & \mbox{ココナッツ・オイル} \\
Coconut milk & \mbox{ココナッツミルク};\allowbreak\ \mbox{ココナッツ・ミルク} \\
Costus spicatus & \mbox{コスツス・スピカツス} \\
Copaiba balm & \mbox{コパイババーム};\allowbreak\ \mbox{コパイバーム};\allowbreak\ \mbox{コパフ・バーム};\allowbreak\ \mbox{ブラジル産コパフバーム} \\
Coriander & \mbox{コリアンダー};\allowbreak\ \mbox{ロシア産コリアンダー} \\
Coriander essence & \mbox{コリアンダー・エッセンス} \\
Coriander seed & \mbox{コリアンダー・シード} \\
Coriander leaf & \mbox{コリアンダーリーフ};\allowbreak\ \mbox{コリアンダー・リーフ} \\
Coffee CO2 extract & \mbox{コーヒー CO2} \\
Coffee absolute & \mbox{コーヒー・アブソリュート};\allowbreak\ \mbox{ブラジリアンコーヒーアブソリュート} \\
Goji berry & \mbox{ゴジベリー（クコの実）} \\
Cypress & \mbox{サイプレス};\allowbreak\ \mbox{イタリアンサイプレス};\allowbreak\ \mbox{イタリアン・サイプレス} \\
Cypress essence & \mbox{サイプレス・エッセンス} \\
Taif rose & \mbox{サウジアラビア産タイフローズ};\allowbreak\ \mbox{タイフローズ};\allowbreak\ \mbox{ターイフ・ローズ} \\
Saffron & \mbox{サフラン};\allowbreak\ \mbox{インディアンサフラン} \\
Sour cherry & \mbox{サワーチェリー};\allowbreak\ \mbox{サワー・チェリー} \\
Sandalwood ORPUR™ & \mbox{サンダルウッドORPUR™} \\
Santos green coffee & \mbox{サントス・グリーンコーヒー} \\
Sambac jasmine & \mbox{サンバック・ジャスミン} \\
Sycamore & \mbox{シカモア（セイヨウカジカエデ）} \\
Citron & \mbox{シシリア産シダー};\allowbreak\ \mbox{シトロン};\allowbreak\ \mbox{イタリア産シトロン};\allowbreak\ \mbox{シトロン（シチリアン・シダー）};\allowbreak\ \mbox{シトロン（セドラ）} \\
Cistus & \mbox{シスタス};\allowbreak\ \mbox{シスタス（ラブダナム）} \\
Cistus absolute & \mbox{シスタス・アブソリュート};\allowbreak\ \mbox{スペイン産シスタスアブソリュート（ラブダナム）} \\
Cistus essence Orpur & \mbox{シスタス・エッセンスOrpur} \\
Labdanum & \mbox{シスタスラブダナム};\allowbreak\ \mbox{シスタス・ラブダナム};\allowbreak\ \mbox{スパニッシュ・ラブダナム};\allowbreak\ \mbox{スペイン産シスタスラブダナム};\allowbreak\ \mbox{スペイン産ラブダナム};\allowbreak\ \mbox{フランス産ラブダナム};\allowbreak\ \mbox{ラブダナム};\allowbreak\ \mbox{フレンチ・ラブダナム};\allowbreak\ \mbox{フレンチラブダナム};\allowbreak\ \mbox{モロッコ産ラブダナム};\allowbreak\ \mbox{ラブダナム（シストローズ）} \\
Cedar & \mbox{シダー};\allowbreak\ \mbox{シダーウッド};\allowbreak\ \mbox{ヴァージニア・シダー};\allowbreak\ \mbox{ヴァージニアシダー};\allowbreak\ \mbox{アトラスシダー};\allowbreak\ \mbox{ヴァージニア産シダー};\allowbreak\ \mbox{ホワイトシダー};\allowbreak\ \mbox{アトラス・シダー};\allowbreak\ \mbox{レッドシダー};\allowbreak\ \mbox{アトラス・シダーウッド};\allowbreak\ \mbox{アトラス産シダーウッド};\allowbreak\ \mbox{アトラス産シダー};\allowbreak\ \mbox{テキサス産シダーウッド};\allowbreak\ \mbox{ブルーシダー};\allowbreak\ \mbox{モロッコ産アトラス・シダー};\allowbreak\ \mbox{モロッコ産シダーウッド};\allowbreak\ \mbox{ヴァージニア・シダーウッド};\allowbreak\ \mbox{ヴァージニア産シダーウッド};\allowbreak\ \mbox{中国産シダーウッド（チャンチン）} \\
Atlas cedarwood Neoabsolut™ & \mbox{シダーウッドアトラスネオアブソルート™} \\
Virginia cedarwood & \mbox{シダーウッド・バージニア} \\
Cedar essence & \mbox{シダー・エッセンス} \\
Cedar heart & \mbox{ヴァージニアシダーウッドハート} \\
Sichuan pepper & \mbox{シチュアン ペッパー} \\
Citrus & \mbox{シトラス};\allowbreak\ \mbox{シトラス類} \\
Citrus fruits & \mbox{イタリアンシトラスフルーツ} \\
Cinnamon & \mbox{シナモン};\allowbreak\ \mbox{ラオス産シナモン};\allowbreak\ \mbox{スリランカ産シナモン};\allowbreak\ \mbox{マダガスカル産シナモン};\allowbreak\ \mbox{中国産シナモン};\allowbreak\ \mbox{セイロン産シナモン} \\
Cypriol & \mbox{シプリオル};\allowbreak\ \mbox{シプリオール};\allowbreak\ \mbox{インド産シプリオール} \\
Siberian pine & \mbox{シベリアン・パイン} \\
Pine & \mbox{シベリア産パイン（松の木）};\allowbreak\ \mbox{パイン};\allowbreak\ \mbox{パイン（松）};\allowbreak\ \mbox{松};\allowbreak\ \mbox{松（パイン）} \\
Siam benzoin & \mbox{シャム・ベンゾイン} \\
Benzoin & \mbox{シャム産ベンゾイン};\allowbreak\ \mbox{タイ産ベンゾイン};\allowbreak\ \mbox{ベンゾイン};\allowbreak\ \mbox{ラオス産ベンゾイン};\allowbreak\ \mbox{ベンゾイン（安息香）};\allowbreak\ \mbox{安息香} \\
Silkwood blossom & \mbox{シルクウッド・ブロッサム} \\
Sea water & \mbox{シーウォーター};\allowbreak\ \mbox{シー・ウォーター} \\
Sea salt & \mbox{シーソルト};\allowbreak\ \mbox{シー・ソルト} \\
Genista lydia & \mbox{ジェニスタ・リディア} \\
Jasmine absolute & \mbox{ジャスミン・アブソリュート};\allowbreak\ \mbox{ジャスミン・サンバック・アブソリュート};\allowbreak\ \mbox{エジプト産ジャスミン・アブソリュート};\allowbreak\ \mbox{ジャスミンアブソリュート};\allowbreak\ \mbox{インド産ジャスミン・サンバック・アブソリュート};\allowbreak\ \mbox{グラース産ジャスミン・アブソリュート};\allowbreak\ \mbox{ジャスミンサンバック・アブソリュート};\allowbreak\ \mbox{インド産ジャスミンサンバック・アブソリュート};\allowbreak\ \mbox{エジプト産ジャスミンアブソリュート};\allowbreak\ \mbox{中国産ジャスミンサンバック・アブソリュート};\allowbreak\ \mbox{インド産ジャスミンアブソリュート（グランディフローラム} \\
Jasmine sambac infusion & \mbox{ジャスミンサンバック・インフュージョン} \\
Jasmine leaf & \mbox{ジャスミン・リーフ} \\
Japanese gardenia & \mbox{ジャパニーズ・ガーデニア} \\
Japanese cherry blossom & \mbox{ジャパニーズ・チェリーブロッサム} \\
Japanese plum & \mbox{ジャパニーズ・プラム} \\
Pimento & \mbox{ジャマイカ産ペッパー};\allowbreak\ \mbox{ピメント} \\
Juniper essence & \mbox{ジュニパー・エッセンス} \\
Juniper berry & \mbox{ジュニパーベリー};\allowbreak\ \mbox{ジュニパー・ベリー} \\
Ginger & \mbox{ジンジャー};\allowbreak\ \mbox{インディアン・ジンジャー};\allowbreak\ \mbox{インドネシア産ジンジャー};\allowbreak\ \mbox{インド産ジンジャー};\allowbreak\ \mbox{マダガスカル産ジンジャー};\allowbreak\ \mbox{タンザニア産ジンジャー};\allowbreak\ \mbox{ナイジェリアン・ジンジャー} \\
Ginger flower & \mbox{ジンジャー・フラワー};\allowbreak\ \mbox{ジンジャーフラワー} \\
Ginger blossom & \mbox{ジンジャー・ブロッサム} \\
Ginger lily & \mbox{ジンジャー・リリー};\allowbreak\ \mbox{ジンジャーリリー} \\
Narcissus & \mbox{スイセン};\allowbreak\ \mbox{水仙} \\
Sweet orange & \mbox{ブラジル産スイートオレンジ};\allowbreak\ \mbox{スウィート・オレンジ};\allowbreak\ \mbox{イタリア産スウィートオレンジ} \\
Sweet almond & \mbox{スウィート・アーモンド} \\
Skin accord & \mbox{スキン・アコード} \\
Himalayan cedar & \mbox{ヒマラヤスギ};\allowbreak\ \mbox{ヒマラヤ杉} \\
Stephanotis & \mbox{ステファノティス（マダガスカル・ジャスミン）} \\
Strawberry leaf & \mbox{ストロベリー・リーフ} \\
Snowdrop & \mbox{スノードロップ};\allowbreak\ \mbox{スノードロップ（待雪草）} \\
Spices & \mbox{スパイス};\allowbreak\ \mbox{スパイス類} \\
Spicy notes & \mbox{スパイスノート};\allowbreak\ \mbox{スパイス・ノート} \\
Spanish moss & \mbox{スパニッシュ・モス} \\
Smoked tea & \mbox{スモークティー};\allowbreak\ \mbox{スモーク・ティー} \\
Common hawthorn & \mbox{セイヨウサンザシ};\allowbreak\ \mbox{ホーソン（セイヨウサンザシ）} \\
Ceylon cinnamon & \mbox{セイロン・シナモン} \\
Ceylon tea & \mbox{セイロン・ティー} \\
Ceylon black tea & \mbox{セイロン・ブラックティー} \\
Sesame & \mbox{セサミ};\allowbreak\ \mbox{セサミ（ゴマ）} \\
Cetalox & \mbox{セタロックス®};\allowbreak\ \mbox{セタロックス®（アンバーグリス）} \\
Cedrat & \mbox{セドラ};\allowbreak\ \mbox{イタリア産セドラ} \\
Bitter orange & \mbox{セビリア産オレンジ};\allowbreak\ \mbox{ビターオレンジ};\allowbreak\ \mbox{ビター・オレンジ};\allowbreak\ \mbox{カラブリア産ビターオレンジ};\allowbreak\ \mbox{イタリア産ビターオレンジ};\allowbreak\ \mbox{シチリア産ビターオレンジ} \\
Serenolide & \mbox{セレノイド（ホワイトムスク）} \\
Centifolia & \mbox{センティフォリアローズ};\allowbreak\ \mbox{グラース産のセンティフォリア・メイローズ};\allowbreak\ \mbox{センティフォリア・ローズ} \\
Centifolia rose absolute & \mbox{センティフォリアローズ・アブソリュート};\allowbreak\ \mbox{グラース産センティフォリアローズ・アブソリュート} \\
St. John flower absolute & \mbox{セントジョンフラワー・アブソリュート（イモーテル）} \\
Sage essence & \mbox{セージ・エッセンス} \\
Bourbon geranium & \mbox{ゼラニウムブルボン};\allowbreak\ \mbox{ゼラニウム・ブルボン} \\
Rose geranium & \mbox{ゼラニウムローズ};\allowbreak\ \mbox{ローズ・ゼラニウム};\allowbreak\ \mbox{ローズゼラニウム} \\
Mandarin orange & \mbox{ソレント産マンダリン};\allowbreak\ \mbox{マンダリン};\allowbreak\ \mbox{シチリア産マンダリン};\allowbreak\ \mbox{イタリアン・マンダリン};\allowbreak\ \mbox{シシリアン・マンダリン};\allowbreak\ \mbox{イタリアンマンダリン};\allowbreak\ \mbox{シチリアン・マンダリン};\allowbreak\ \mbox{インディアン・マンダリン};\allowbreak\ \mbox{シシリアンマンダリン};\allowbreak\ \mbox{マンダリン・オレンジ};\allowbreak\ \mbox{マンダリンオレンジ};\allowbreak\ \mbox{シチリア産マンダリン・オレンジ};\allowbreak\ \mbox{シチリア産マンダリンオレンジ};\allowbreak\ \mbox{カラブリア産マンダリンオレンジ};\allowbreak\ \mbox{シシリー産マンダリン・オレンジ};\allowbreak\ \mbox{チュニジア産マンダリン・オレンジ} \\
Solar wood & \mbox{ソーラー・ウッド} \\
Thyme & \mbox{タイム};\allowbreak\ \mbox{モロッコ産タイム} \\
Tagetes & \mbox{タゲット（メキシカン・マリーゴールド）} \\
Tangerine & \mbox{タジェリン};\allowbreak\ \mbox{タンジェリン};\allowbreak\ \mbox{イタリアン・タンジェリン};\allowbreak\ \mbox{カラブリアン・タンジェリン} \\
Tuscan iris & \mbox{タスカン・アイリス} \\
Thanaka wood & \mbox{タナカウッド};\allowbreak\ \mbox{タナカ・ウッド} \\
Mimosa & \mbox{タヌロン産ミモザ};\allowbreak\ \mbox{プロヴァバンス産ミモザ};\allowbreak\ \mbox{プロヴァンス産ミモザ};\allowbreak\ \mbox{ミモザ};\allowbreak\ \mbox{インド産ミモザ};\allowbreak\ \mbox{グラース・ミモザ};\allowbreak\ \mbox{グラース産ミモザ} \\
Tobacco absolute & \mbox{タバコ・アブソリュート} \\
Tobacco leaf & \mbox{タバコリーフ};\allowbreak\ \mbox{タバコ・リーフ} \\
Tarragon & \mbox{タラゴン};\allowbreak\ \mbox{タラゴン（ヨモギ）} \\
Turkish red rose & \mbox{ターキッシュ・レッドローズ} \\
Turkish red rose absolute & \mbox{ターキッシュ・レッドローズ・アブソリュート} \\
Turkish red rose petal & \mbox{ターキッシュ・レッドローズ・ペタル} \\
Turkish rose & \mbox{ターキッシュローズ};\allowbreak\ \mbox{ターキッシュ・ローズ};\allowbreak\ \mbox{ターキッシュローズ Orpur®} \\
Turmeric & \mbox{ターメリック};\allowbreak\ \mbox{ラオス産ターメリック} \\
Diamond orchid & \mbox{ブラジル産ダイヤモンドオーキッド} \\
Damask rose & \mbox{ダマスカスローズ};\allowbreak\ \mbox{ダマスクローズ};\allowbreak\ \mbox{ダマスク・ローズ};\allowbreak\ \mbox{トルコ産ダマスクローズ};\allowbreak\ \mbox{モロッコ産ダマスクローズ} \\
Prune absolute & \mbox{ダマスカス産プルーンアブソリュート} \\
Damask pink rose & \mbox{ダマスク・ピンクローズ} \\
Damask rose absolute & \mbox{ダマスクローズアブソリュート};\allowbreak\ \mbox{ダマスクローズ・アブソリュート} \\
Damask rose essence & \mbox{ダマスクローズ・エッセンス} \\
Dark chocolate & \mbox{ダークチョコレート};\allowbreak\ \mbox{ダーク・チョコレート} \\
Cherry blossom & \mbox{チェリー・ブロッサム};\allowbreak\ \mbox{チェリーブロッサム} \\
Chinese orange & \mbox{チャイニーズ・オレンジ} \\
Chinese cedar & \mbox{チャイニーズ・シダー} \\
Chinese cinnamon wood & \mbox{チャイニーズ・シナモンウッド} \\
Chinese peony & \mbox{チャイニーズ・ピオニー（雲南省）} \\
Chinese pink peony & \mbox{チャイニーズ・ピンクピオニー} \\
Champaca & \mbox{インドネシア産チャンパカ};\allowbreak\ \mbox{チャンパカ} \\
Tuberose absolute & \mbox{インド産チュベローズアブソリュート};\allowbreak\ \mbox{チュベローズ・アブソリュート};\allowbreak\ \mbox{チューベローズ・アブソリュート};\allowbreak\ \mbox{インド産チューベローズ・アブソリュート};\allowbreak\ \mbox{チューベローズアブソリュート};\allowbreak\ \mbox{チューベローズ・アブソリュー} \\
Tuberose & \mbox{チューベローズ};\allowbreak\ \mbox{インド産チューベローズ};\allowbreak\ \mbox{インディアン・チューベローズ} \\
Datura & \mbox{チョウセンアサガオ（ダチュラ）};\allowbreak\ \mbox{チョウセンアサガオ} \\
Tiare flower & \mbox{ティアラフラワー};\allowbreak\ \mbox{タヒチ産ティアラフラワー};\allowbreak\ \mbox{ティアラ・フラワー};\allowbreak\ \mbox{ティアレフラワー};\allowbreak\ \mbox{ティアレ・フラワー} \\
Tiare flower accord & \mbox{ティアレ・フラワー・アコード} \\
Timut pepper & \mbox{ティムットペッパー};\allowbreak\ \mbox{ティムット・ペッパー} \\
Diamante citron & \mbox{ディアマンテ・シトロン} \\
Dates & \mbox{デーツ};\allowbreak\ \mbox{デーツ（ナツメヤシの実）} \\
Violet & \mbox{トゥールーズ産ヴァイオレット};\allowbreak\ \mbox{バイオレット};\allowbreak\ \mbox{バイオレット（すみれ）};\allowbreak\ \mbox{パルマ産ヴァイオレット};\allowbreak\ \mbox{ヴァイオレット};\allowbreak\ \mbox{ヴィオレット} \\
Toffee & \mbox{トフィー};\allowbreak\ \mbox{トフィー（塩キャラメル）} \\
Toffee apple & \mbox{トフィー アップル};\allowbreak\ \mbox{トフィーアップル} \\
Toffee caramel & \mbox{トフィー・キャラメル} \\
Pittosporum & \mbox{トベラ};\allowbreak\ \mbox{ピットスポルム（トベラ）} \\
Tomato leaf & \mbox{トマトリーフ};\allowbreak\ \mbox{トマト・リーフ} \\
Tropical fruits & \mbox{トロピカル・フルーツ} \\
Tonka bean & \mbox{トンカ};\allowbreak\ \mbox{トンカビーン};\allowbreak\ \mbox{ブラジル産トンカビーン};\allowbreak\ \mbox{トンカ・ビーン};\allowbreak\ \mbox{ペルー産トンカビーン};\allowbreak\ \mbox{トンカビーンズ};\allowbreak\ \mbox{ベネズエラ産トンカビーン} \\
Tonka bean absolute & \mbox{トンカビーン・アブソリュート} \\
Roasted tonka absolute & \mbox{トンカ・ローステッド・アブソリュート} \\
St. Thomas bay leaf & \mbox{ドミニカ共和国産セントトーマス・ベイ・リーフ} \\
Driftwood & \mbox{ドラフトウッド};\allowbreak\ \mbox{ドリフトウッド};\allowbreak\ \mbox{ドリフトウッド（流木）} \\
Dreamwood & \mbox{ドリームウッド®} \\
Night-blooming jasmine & \mbox{ナイト・ブルーミング・ジャスミン} \\
Natural musk & \mbox{ナチュラル・ムスク} \\
Nutmeg & \mbox{ナツメグ};\allowbreak\ \mbox{インドネシアン・ナツメグ};\allowbreak\ \mbox{ジャワ産ナツメグ} \\
Nutmeg flower & \mbox{ナツメグフラワー};\allowbreak\ \mbox{ナツメグ・フラワー} \\
Narcissus absolute & \mbox{ナルシス・アブソリュート} \\
Black locust & \mbox{ニシアカシア};\allowbreak\ \mbox{ニセアカシア} \\
Nepal pepper & \mbox{ネパールペッパー} \\
Neroli & \mbox{ネロリ};\allowbreak\ \mbox{チュニジア産ネロリ};\allowbreak\ \mbox{チュニジアン・ネロリ};\allowbreak\ \mbox{グラース産ネロリ};\allowbreak\ \mbox{ネロリビガラード};\allowbreak\ \mbox{アフリカ産ネロリ};\allowbreak\ \mbox{イタリアンネロリ};\allowbreak\ \mbox{エジプト産ネロリ};\allowbreak\ \mbox{カラブリア産ネロリ};\allowbreak\ \mbox{ネロリアコード};\allowbreak\ \mbox{ネロリペタル・アコード};\allowbreak\ \mbox{プロヴァンス産ネロリ};\allowbreak\ \mbox{ホワイトネロリ};\allowbreak\ \mbox{モロッコ産ネロリ} \\
Neroli essence & \mbox{セビリア・オレンジ・ネロリ・エッセンス};\allowbreak\ \mbox{チュニジア産ネロリエッセンス} \\
Neroli absolute & \mbox{ネロリ・アブソリュート} \\
Neroli oil & \mbox{モロッカンネロリオイル} \\
Hibiscus seed & \mbox{ハイビスカスシード};\allowbreak\ \mbox{ハイビスカス・シード} \\
Hibiscus blossom & \mbox{ハイビスカス・ブロッサム} \\
Violet leaf & \mbox{バイオレット・リーフ};\allowbreak\ \mbox{バイオレットリーフ};\allowbreak\ \mbox{パルマ産ヴァイオレットリーブス};\allowbreak\ \mbox{ヴァイオレットの葉};\allowbreak\ \mbox{ヴァイオレット・リーフ};\allowbreak\ \mbox{ヴァイオレットリーフ};\allowbreak\ \mbox{エジプト産ヴァイオレット・リーフ};\allowbreak\ \mbox{グラース産ヴァイオレット・リーフ} \\
Basil & \mbox{バジル};\allowbreak\ \mbox{エジプト産バジル} \\
Banana flower & \mbox{バナナ・フラワー} \\
Banana leaf & \mbox{バナナ・リーフ} \\
Vanilla CO2 extract & \mbox{バニラCO2};\allowbreak\ \mbox{バーボンバニラCO2} \\
Vanilla absolute & \mbox{バニラアブソリュート};\allowbreak\ \mbox{バニラ・アブソリュート};\allowbreak\ \mbox{マダガスカル産バニラ・アブソリュート};\allowbreak\ \mbox{タヒチ産バニラ・アブソリュート};\allowbreak\ \mbox{バーボン・バニラ・アブソリュート};\allowbreak\ \mbox{ブルボンバニラ・アブソリュート};\allowbreak\ \mbox{パプアニューギニア産バニラアブソリュート} \\
Vanilla orchid & \mbox{バニラオーキッド};\allowbreak\ \mbox{バニラ・オーキッド} \\
Vanilla custard & \mbox{バニラ・カスタード} \\
Vanilla beans & \mbox{タヒチ産バニラビーンズ};\allowbreak\ \mbox{バニラビーンズ} \\
Balsamic resin & \mbox{バルサミック・レジン} \\
Peru balsam & \mbox{ペルーバルサム};\allowbreak\ \mbox{ペルー・バルサム};\allowbreak\ \mbox{ペル・バルサム} \\
Balsam & \mbox{エジプト産バルサム};\allowbreak\ \mbox{ペルー産バルサム} \\
Balsam fir & \mbox{バルサム・ファー} \\
Valencia orange & \mbox{バレンシア・オレンジ} \\
Bamboo & \mbox{バンブー};\allowbreak\ \mbox{バンブー（竹）} \\
Virginia cedarwood & \mbox{バージニア・シダーウッド} \\
Birch & \mbox{バーチ};\allowbreak\ \mbox{バーチ（白樺）} \\
Verbena & \mbox{バーベナ};\allowbreak\ \mbox{ヴァーベナ} \\
Vermont red maple & \mbox{バーモント・レッドメープル} \\
Pineapple leaf & \mbox{パイナップル・リーフ} \\
Pine tree & \mbox{パインツリー};\allowbreak\ \mbox{パインツリー（松）} \\
Pine tree needle & \mbox{パイン・ツリー・ニードル} \\
Powdery notes & \mbox{パウダリー・ノーツ};\allowbreak\ \mbox{パウダリーノート};\allowbreak\ \mbox{パウダリー・ノート} \\
Patchouli & \mbox{パチョリ};\allowbreak\ \mbox{インドネシア産パチョリ};\allowbreak\ \mbox{インド産パチョリ};\allowbreak\ \mbox{パチュリ};\allowbreak\ \mbox{インドシナ産パチョリ};\allowbreak\ \mbox{ジャワ産パチョリ};\allowbreak\ \mbox{ヴァージニア産パチョリ} \\
Patchouli absolute & \mbox{パチョリ・アブソリュート} \\
Patchouli essence & \mbox{インドネシア産パチョリエッセンス};\allowbreak\ \mbox{パチョリエッセンス} \\
Patchouli heart & \mbox{パチョリハート};\allowbreak\ \mbox{インドネシア産パチョリハート} \\
Patchouli fraction & \mbox{パチョリ・フラクション} \\
Patchouli leaf & \mbox{インドネシア産パチョリ・リーフ};\allowbreak\ \mbox{インドネシアン・パチョリ・リーフ};\allowbreak\ \mbox{インドネシアン・パチョリリーフ};\allowbreak\ \mbox{インドネシア産パチョリリーフ};\allowbreak\ \mbox{パチョリリーフ} \\
Passionflower & \mbox{パッションフラワー};\allowbreak\ \mbox{パッション・フラワー} \\
Papyrus & \mbox{パピルス};\allowbreak\ \mbox{インド産パピルス} \\
Paradisone & \mbox{パラディゾン®};\allowbreak\ \mbox{パラディゾン};\allowbreak\ \mbox{パラディゾーネ} \\
Persimmon & \mbox{パーシモン（柿）} \\
Purple sage & \mbox{パープル・セージ} \\
Hyacinth & \mbox{ヒアシンス};\allowbreak\ \mbox{ヒヤシンス} \\
Bigarade & \mbox{ビガラード（ダイダイ）} \\
Bitter almond & \mbox{ビターアーモンド};\allowbreak\ \mbox{ビター・アーモンド} \\
Bitter orange essence & \mbox{コートジボワール産ビターオレンジエッセンス} \\
Bitter cacao & \mbox{ビター・カカオ} \\
Large strawberry & \mbox{ビッグ・ストロベリー};\allowbreak\ \mbox{イタリア産ビッグ・ストロベリー};\allowbreak\ \mbox{ビッグストロベリー} \\
Beeswax & \mbox{ビーズワックス};\allowbreak\ \mbox{ビーズワックス（蜜蝋）};\allowbreak\ \mbox{蜜蝋} \\
Pear blossom & \mbox{ピアー・ブロッサム};\allowbreak\ \mbox{ペアー・ブロッサム} \\
Alba truffle & \mbox{ピエモンテ産アルバトリュフ} \\
Pimento accord & \mbox{ピメント・アコード} \\
Pimento berry & \mbox{ピメントベリー};\allowbreak\ \mbox{ピメントベリー（オールスパイス）} \\
Pimento leaf & \mbox{ピメント・リーフ} \\
Pink grapefruit & \mbox{ピンク・グレープフルーツ};\allowbreak\ \mbox{ピンクグレープフルーツ} \\
Pink tulip & \mbox{ピンク・チューリップ} \\
Pink pepper & \mbox{ピンクペッパー};\allowbreak\ \mbox{ピンク・ペッパー};\allowbreak\ \mbox{レユニオン島産ピンクペッパー};\allowbreak\ \mbox{ブラジル産ピンクペッパー};\allowbreak\ \mbox{マダガスカル産ピンクペッパー} \\
Pink pepper essence & \mbox{ピンクペッパー・エッセンス} \\
Pink pepper essence CO2 & \mbox{マダガスカル産ピンクペッパーエッセンスCO2} \\
Pink rose & \mbox{ピンクローズ};\allowbreak\ \mbox{ピンク・ローズ} \\
Pink violet & \mbox{ピンク・ヴァイオレット} \\
Peach & \mbox{ピーチ};\allowbreak\ \mbox{シチリア産ピーチ} \\
Peach blossom & \mbox{ピーチブロッサム};\allowbreak\ \mbox{ピーチ・ブロッサム} \\
Fir balsam absolute & \mbox{ファーバルサム・アブソリュート} \\
Fig tree & \mbox{フィグツリー};\allowbreak\ \mbox{フィグ・ツリー} \\
Fig nectar & \mbox{フィグ・ネクター} \\
Fig leaf & \mbox{フィグリーフ};\allowbreak\ \mbox{フィグ・リーフ} \\
Fig leaf accord & \mbox{フィグリーフ・アコード} \\
Physcool & \mbox{フィスクール®} \\
Firabs iris & \mbox{フィラブス・アイリス} \\
Fennel seed & \mbox{フェンネルシード（ウイキョウ）} \\
Flower petal & \mbox{フラワー・ペタル} \\
Flower petals & \mbox{フラワー・ペタルズ} \\
Frankincense absolute & \mbox{フランキンセンス・アブソリュート} \\
Frangipani & \mbox{フランジパニ};\allowbreak\ \mbox{フランジパニ（プルメリア）};\allowbreak\ \mbox{プルメリア} \\
Freesia & \mbox{フリージア};\allowbreak\ \mbox{南アフリカ産フリージア} \\
Fruity notes & \mbox{フルーティ・ノーツ};\allowbreak\ \mbox{フルーティノート};\allowbreak\ \mbox{フルーティ・ノート};\allowbreak\ \mbox{フルーティーノート} \\
Fleur de sel & \mbox{フルール・ド・セル（天日塩）} \\
Fresh spices & \mbox{フレッシュ・スパイス} \\
Frosted orange & \mbox{シチリア産フロストオレンジ} \\
Frozen grapefruit & \mbox{フローズン・グレープフルーツ} \\
Floral notes & \mbox{フローラルノート};\allowbreak\ \mbox{フローラル・ノート} \\
Buchu & \mbox{ブチュ};\allowbreak\ \mbox{ブチュー} \\
Brown sugar & \mbox{ブラウンシュガー};\allowbreak\ \mbox{ブラウン・シュガー} \\
Black amber & \mbox{ブラックアンバー};\allowbreak\ \mbox{ブラック・アンバー} \\
Black elder & \mbox{ブラックエルダー};\allowbreak\ \mbox{ブラック・エルダー} \\
Black orchid & \mbox{ブラックオーキッド};\allowbreak\ \mbox{ブラック・オーキッド} \\
Blackcurrant accord & \mbox{ブラックカラント・アコード} \\
Blackcurrant absolute & \mbox{ブラックカラント・アブソルート} \\
Blackcurrant bud & \mbox{ブラックカラント・バッド};\allowbreak\ \mbox{ブラックカラントバッド} \\
Blackcurrant bud absolute & \mbox{ブラックカラントバッド・アブソリュート};\allowbreak\ \mbox{ブラックカラント・バッド・アブソリュート} \\
Blackcurrant blossom & \mbox{ブラックカラント・ブロッサム};\allowbreak\ \mbox{ブルゴーニュ産ブラックカレントブロッサム} \\
Blackcurrant liqueur & \mbox{ブラックカラント・リキュール} \\
Blackcurrant leaf & \mbox{ブラックカラントリーフ};\allowbreak\ \mbox{ブラックカラント・リーフ} \\
Black cardamom & \mbox{ブラックカルダモン};\allowbreak\ \mbox{ブラック・カルダモン} \\
Black cedarwood & \mbox{ブラック・シダーウッド} \\
Black tea & \mbox{ブラックティー};\allowbreak\ \mbox{インド産ブラックティー};\allowbreak\ \mbox{スリランカ産ブラックティー} \\
Black truffle & \mbox{ブラックトリュフ};\allowbreak\ \mbox{ブラック・トリュフ} \\
Black violet & \mbox{ブラックバイオレット};\allowbreak\ \mbox{ブラック・ヴァイオレット} \\
Black vanilla & \mbox{ブラック・バニラ};\allowbreak\ \mbox{ブラックバニラ} \\
Black lilac & \mbox{ブラック・ライラック} \\
Black licorice & \mbox{ブラックリコリス};\allowbreak\ \mbox{ブラック・リコリス} \\
Black leather accord & \mbox{ブラックレザー・アコード} \\
Blood orange & \mbox{ブラッド・オレンジ};\allowbreak\ \mbox{ブラッドオレンジ};\allowbreak\ \mbox{シチリア産ブラッドオレンジ} \\
Blood mandarin & \mbox{ブラッド・マンダリン} \\
Vulcanolide & \mbox{ブルカノライド®（ミネラルムスク）} \\
Bulgarian rose & \mbox{ブルガリアン・ローズ};\allowbreak\ \mbox{ブルガリアンローズ};\allowbreak\ \mbox{ブルガリアン･ローズ};\allowbreak\ \mbox{ブルガリアンローズ Orpur®};\allowbreak\ \mbox{ブルガリアンローズ（コンクリート＆オイル）};\allowbreak\ \mbox{ブルガリンローズ} \\
Bulgarian rose essence & \mbox{ブルガリアン・ローズ・エッセンス};\allowbreak\ \mbox{ブルガリアンローズ・エッセンス} \\
Bulgarian rose absolute & \mbox{ブルガリアンローズ・アブソリュート};\allowbreak\ \mbox{ブルガリアンローズアブソリュート} \\
Blue chamomile & \mbox{ブルーカモミール};\allowbreak\ \mbox{ブルー・カモミール};\allowbreak\ \mbox{モロッコ産ブルーカモミール} \\
Blue spruce & \mbox{ブルー・スプルース} \\
Blue sequoia & \mbox{ブルー セコイア};\allowbreak\ \mbox{ブルーセコイア} \\
Blue lotus & \mbox{ブルー・ロータス} \\
Brogiotto black fig & \mbox{ブロジオット・ブラック・フィグ} \\
Petitgrain & \mbox{プチグレイン};\allowbreak\ \mbox{プチグレン};\allowbreak\ \mbox{パラグアイ産プチグレン};\allowbreak\ \mbox{イタリア産プチグレン};\allowbreak\ \mbox{シチリア産プチグレン};\allowbreak\ \mbox{プチグレンレモン};\allowbreak\ \mbox{レモンプチグレン};\allowbreak\ \mbox{プチブレン} \\
Plum & \mbox{プラム};\allowbreak\ \mbox{マダガスカル産プラム} \\
Plum wood & \mbox{プラムウッド};\allowbreak\ \mbox{プラム・ウッド} \\
Plum blossom & \mbox{プラム・ブロッサム} \\
Primofiore lemon & \mbox{プリモフィオーレ・レモン} \\
Precious woods & \mbox{プレシャスウッズ};\allowbreak\ \mbox{プレシャスウッド} \\
Prosecco accord & \mbox{プロセッコ・アコード} \\
Lavender & \mbox{プロバンス産ラベンダー};\allowbreak\ \mbox{プロヴァンス産ラベンダー};\allowbreak\ \mbox{ラベンダー};\allowbreak\ \mbox{フレンチラベンダー};\allowbreak\ \mbox{フレンチ・ラベンダー};\allowbreak\ \mbox{ブルガリア産ラベンダー} \\
Lavandin & \mbox{プロヴァンス産ラヴァンディン} \\
Pu-erh tea & \mbox{プーアル・ティー} \\
Hay & \mbox{ヘイ（干し草）};\allowbreak\ \mbox{干し草};\allowbreak\ \mbox{干草} \\
Heather accord & \mbox{ヘザー・アコード} \\
Helvetolide & \mbox{ヘルヴェトライド®};\allowbreak\ \mbox{ヘルヴェトライド} \\
Hazel blossom & \mbox{ヘーゼル・ブロッサム} \\
Bay leaf & \mbox{ベイリーフ};\allowbreak\ \mbox{ベイリーフ（ローリエ）};\allowbreak\ \mbox{ローリエ} \\
Vetiver essence & \mbox{ハイチ産ベチバーエッセンス};\allowbreak\ \mbox{ハイチ産ベチバー・エッセンス} \\
Vetiver oil & \mbox{ジャワ産ベチバーオイル};\allowbreak\ \mbox{ベチバーオイル} \\
Vetiver concentrate & \mbox{ベチバー・コンセントレイト} \\
Bergamot essence & \mbox{ベルガモット・エッセンス};\allowbreak\ \mbox{カラブリアン・ベルガモット・エッセンス};\allowbreak\ \mbox{ベルガモットエッセンス} \\
Bergamot leaf & \mbox{ベルガモット・リーフ} \\
Velvety rose & \mbox{ベルベッティ・ローズ} \\
Benzoin essence & \mbox{ベンゾイン・エッセンス} \\
Pear & \mbox{ペアー};\allowbreak\ \mbox{洋ナシ};\allowbreak\ \mbox{洋梨} \\
Petalia & \mbox{ペタリア};\allowbreak\ \mbox{ペタリア®} \\
Pepper & \mbox{ブラックペッパー};\allowbreak\ \mbox{ペッパー};\allowbreak\ \mbox{ブラック・ペッパー};\allowbreak\ \mbox{マダガスカル産ブラックペッパー};\allowbreak\ \mbox{インド産ブラックペッパー};\allowbreak\ \mbox{ブルボンペッパー};\allowbreak\ \mbox{胡椒} \\
Peppermint & \mbox{ペパーミント};\allowbreak\ \mbox{ブラジリアン・ペパーミント} \\
Whipped cream & \mbox{ホイップクリーム};\allowbreak\ \mbox{ホイップドクリーム} \\
Hot iron & \mbox{ホット・アイロン} \\
White iris & \mbox{ホワイトイリス};\allowbreak\ \mbox{ホワイト・アイリス} \\
White amber & \mbox{ホワイトアンバー};\allowbreak\ \mbox{ホワイト・アンバー} \\
White wood & \mbox{ホワイトウッズ};\allowbreak\ \mbox{ホワイト・ウッズ};\allowbreak\ \mbox{ホワイトウッド};\allowbreak\ \mbox{ホワイト・ウッド} \\
White wood accord & \mbox{ホワイトウッド・アコード} \\
White orchid & \mbox{ホワイト・オーキッド} \\
White camellia & \mbox{ホワイト・カメリア} \\
White currant & \mbox{ホワイト・カラント} \\
White oleander & \mbox{ホワイト・キョウチクトウ} \\
White sandalwood & \mbox{ホワイト・サンダルウッド};\allowbreak\ \mbox{ホワイトサンダルウッド} \\
White cedar extract & \mbox{ホワイト・シダー・エキス} \\
White jasmine & \mbox{モロッコ産ホワイト・ジャスミン} \\
White ginger lily & \mbox{ホワイト・ジンジャー・リリー} \\
White suede & \mbox{ホワイトスエード};\allowbreak\ \mbox{ホワイト・スエード} \\
White tobacco & \mbox{ホワイトタバコ};\allowbreak\ \mbox{ホワイト・タバコ} \\
White chocolate & \mbox{ホワイトチョコレート};\allowbreak\ \mbox{ホワイト・チョコレート} \\
White tea & \mbox{ホワイトティー};\allowbreak\ \mbox{ホワイト・ティー} \\
White honey & \mbox{ホワイトハニー};\allowbreak\ \mbox{ホワイト・ハニー} \\
White hyacinth & \mbox{ホワイト・ヒヤシンス} \\
White peach & \mbox{ホワイトピーチ};\allowbreak\ \mbox{ホワイト・ピーチ} \\
White flowers & \mbox{ホワイトフラワーズ};\allowbreak\ \mbox{ホワイト・フラワーズ} \\
White pepper & \mbox{ホワイトペッパー};\allowbreak\ \mbox{ホワイト・ペッパー} \\
White rose & \mbox{ホワイトローズ};\allowbreak\ \mbox{ホワイト・ローズ} \\
Bondi Beach flower & \mbox{ボンダイビーチ・フラワー} \\
Pomelo & \mbox{ポメロ};\allowbreak\ \mbox{ポメロ（ザボン）} \\
Mountain air & \mbox{マウンテン・エアー} \\
Macassar ebony wood & \mbox{マカッサル・エボニーウッド} \\
Magnolia & \mbox{マグノリア};\allowbreak\ \mbox{中国産マグノリア};\allowbreak\ \mbox{インド産マグノリア} \\
Magnolia absolute & \mbox{マグノリア・アブソリュート} \\
Marjoram & \mbox{マジョラム};\allowbreak\ \mbox{マジョラム（シソ科）} \\
Macassar ebony & \mbox{マッカサル・エボニー} \\
Mayotte vanilla & \mbox{マヨット・バニラ} \\
Marigold & \mbox{マリゴールド};\allowbreak\ \mbox{マリーゴールド};\allowbreak\ \mbox{アフリカンマリーゴールド} \\
Rum & \mbox{マルティニーク産ラム};\allowbreak\ \mbox{ラム} \\
Quince & \mbox{マルメロ};\allowbreak\ \mbox{マルメロ（西洋かりん）} \\
Mango & \mbox{マンゴ};\allowbreak\ \mbox{マンゴー} \\
Mango blossom & \mbox{マンゴー・ブロッサム} \\
Mandarin essence & \mbox{マンダリン・エッセンス} \\
Mandarin orange peel & \mbox{マンダリンオレンジの皮} \\
Mandarin orange leaf & \mbox{マンダリン・オレンジ・リーフ} \\
Mandarin sorbet & \mbox{マンダリン・シャーベット} \\
Mandarin blossom & \mbox{マンダリン・ブロッサム} \\
Mandarin leaf & \mbox{マンダリン・リーフ} \\
Myrtle & \mbox{マートル};\allowbreak\ \mbox{マートル（ギンバイカ）} \\
Mimosa absolute & \mbox{ミモザアブソリュート};\allowbreak\ \mbox{ミモザ・アブソリュート} \\
Mimosa blossom & \mbox{ミモザ・ブロッサム} \\
Muguet aldehyde & \mbox{ミュゲ・アルデハイド} \\
Mirabelle & \mbox{ミラベル};\allowbreak\ \mbox{ミラベル（西洋スモモ）} \\
Milk chocolate & \mbox{ミルク・チョコレート} \\
Myrrh & \mbox{ミルラ};\allowbreak\ \mbox{ミルラ（没薬）};\allowbreak\ \mbox{ソマリア産ミルラ};\allowbreak\ \mbox{没薬};\allowbreak\ \mbox{没薬（ミルラ）} \\
Myrrh ORPUR™ & \mbox{ミルラORPUR™} \\
Mint & \mbox{ミント};\allowbreak\ \mbox{イタリアンミント} \\
Musky blond wood & \mbox{ムスキーブロンドウッド};\allowbreak\ \mbox{ムスキー・ブロンドウッド} \\
Musk & \mbox{ムスク};\allowbreak\ \mbox{ホワイトムスク};\allowbreak\ \mbox{ホワイト・ムスク};\allowbreak\ \mbox{ブラックムスク};\allowbreak\ \mbox{クリスタルムスク};\allowbreak\ \mbox{ベルベットムスク};\allowbreak\ \mbox{ムスク（エキサルトン）} \\
Musk vodka & \mbox{ムスク・ウォッカ} \\
May rose & \mbox{メイローズ};\allowbreak\ \mbox{グラース産メイローズ};\allowbreak\ \mbox{ローズドゥメ};\allowbreak\ \mbox{ローズ・ドゥメ};\allowbreak\ \mbox{ローズドメ};\allowbreak\ \mbox{ローズドメイ} \\
May rose absolute & \mbox{メイローズ・アブソリュート};\allowbreak\ \mbox{グラース産メイローズ・アブソリュート};\allowbreak\ \mbox{メイローズアブソリュート};\allowbreak\ \mbox{グラース産ローズドゥメ・アブソリュート};\allowbreak\ \mbox{ローズドゥメ・アブソリュート} \\
Mexican vanilla & \mbox{メキシカン・バニラ} \\
Lime & \mbox{メキシコ産ライム};\allowbreak\ \mbox{ライム} \\
Metallic notes & \mbox{メタリックノート};\allowbreak\ \mbox{メタリック・ノート} \\
Methyl ionone & \mbox{メチルイオノン（＝ヴァイオレット）} \\
Methyl pamplemousse & \mbox{メチル・パンプルムース} \\
Melon & \mbox{メロン};\allowbreak\ \mbox{メロン（ヘリオナール）} \\
Mace & \mbox{メース};\allowbreak\ \mbox{メース（ナツメグの皮）} \\
Moss & \mbox{モス};\allowbreak\ \mbox{モス（苔）} \\
Mock orange & \mbox{モック・オレンジ} \\
Mugwort & \mbox{ヨモギ（アルテミシア）} \\
Lychee & \mbox{ライチ};\allowbreak\ \mbox{マダガスカル産ライチ} \\
Lime blossom & \mbox{ライム・ブロッサム} \\
Raspberry bloom & \mbox{ラズベリー・ブルーム} \\
Raspberry leaf & \mbox{ラズベリー・リーフ} \\
Lava stone & \mbox{ラバストーン（溶岩石）} \\
Lavender absolute & \mbox{ラベンダー・アブソリュート} \\
Lavender extract & \mbox{ラベンダー・エクストラクト} \\
Litsea cubeba & \mbox{インドネシア産リツアクベバ} \\
Flax flower & \mbox{リネンフラワー（亜麻）} \\
Lily & \mbox{リリー};\allowbreak\ \mbox{百合} \\
Linden blossom & \mbox{リンデン・ブロッサム};\allowbreak\ \mbox{リンデンブロッサム};\allowbreak\ \mbox{菩提樹（リンデン・ブロッサム）} \\
Rhubarb leaf & \mbox{ルバーブ・リーフ} \\
Leather accord & \mbox{レザーアコード};\allowbreak\ \mbox{レザー・アコード} \\
Redcurrant & \mbox{レッドカラント};\allowbreak\ \mbox{レッド・カラント} \\
Redcurrant blossom & \mbox{レッドカラント・ブロッサム} \\
Red sandalwood & \mbox{レッド・サンダルウッド} \\
Red spider lily & \mbox{レッド・スパイダーリリー} \\
Red thyme & \mbox{レッド・タイム} \\
Red peony & \mbox{レッド・ピオニー};\allowbreak\ \mbox{レッドピオニー} \\
Red peach & \mbox{レッド・ピーチ} \\
Red fruits & \mbox{レッドフルーツ};\allowbreak\ \mbox{レッド・フルーツ} \\
Red berries & \mbox{レッドベリー};\allowbreak\ \mbox{レッド・ベリー};\allowbreak\ \mbox{レッドベリーズ};\allowbreak\ \mbox{レッド・ベリーズ} \\
Red lily & \mbox{レッド・リリー} \\
Red rose & \mbox{レッドローズ};\allowbreak\ \mbox{レッド・ローズ} \\
Lady of the night & \mbox{レディ・オブ・ザ・ナイト（月下美人）} \\
Lemon zest & \mbox{レモン・ゼスト};\allowbreak\ \mbox{イタリア産レモンゼスト};\allowbreak\ \mbox{レモンゼスト} \\
Lemon geranium & \mbox{レモンゼラニウム（ペラルゴニウム）} \\
Lemon verbena & \mbox{レモン・バーベナ};\allowbreak\ \mbox{レモンバーベナ};\allowbreak\ \mbox{レモンヴァーベナ} \\
Lemon blossom & \mbox{レモン・ブロッサム};\allowbreak\ \mbox{レモンブロッサム} \\
Mastic absolute & \mbox{レンティスコ(マスティック) アブソリュート} \\
Rosa damascena & \mbox{ロサ・ダマスケナ} \\
Rosa rubiginosa & \mbox{ロサ・ルビギノーサ} \\
Roasted oak & \mbox{ローステッドオーク};\allowbreak\ \mbox{ローステッド・オーク} \\
Roasted coffee bean & \mbox{ローステッド・コーヒービーン} \\
Roasted coffee beans & \mbox{ローステッド・コーヒー・ビーンズ} \\
Rose & \mbox{ローズ};\allowbreak\ \mbox{モロッコ産ローズ};\allowbreak\ \mbox{グラースローズ};\allowbreak\ \mbox{トルコ産ローズ};\allowbreak\ \mbox{グラース・ローズ};\allowbreak\ \mbox{ブルガリア産ローズ};\allowbreak\ \mbox{エジプト産ローズ};\allowbreak\ \mbox{グラース産ローズ};\allowbreak\ \mbox{ブラジル産ローズ} \\
Rose and green accord & \mbox{ローズ＆グリーンアコード} \\
Rose CO2 extract & \mbox{グラースローズCO2} \\
Rose absolute & \mbox{ローズアブソリュート};\allowbreak\ \mbox{ローズ・アブソリュート} \\
Rose water & \mbox{ローズウォーター};\allowbreak\ \mbox{ローズ・ウォーター} \\
Rosewood & \mbox{ブラジリアン・ローズウッド};\allowbreak\ \mbox{ブラジル産ローズウッド};\allowbreak\ \mbox{パリサンダー・ローズウッド};\allowbreak\ \mbox{ローズウッド};\allowbreak\ \mbox{パリサンダーローズウッド} \\
Rose essential & \mbox{トルコ産ローズエッセンシャル} \\
Rose essential LMR & \mbox{ローズ・エッセンシャルLMR} \\
Rose essence & \mbox{ブルガリア産ローズエッセンス};\allowbreak\ \mbox{ローズエッセンス};\allowbreak\ \mbox{ブルガリア産ローズ・エッセンス} \\
Rose centifolia & \mbox{グラース産ローズセンチフォリア} \\
Rose Turkish delight & \mbox{ローズ・ターキッシュ・デライト（ロクム）} \\
Rose tincture & \mbox{ローズ・チンキ} \\
Rose Neoabsolut & \mbox{ローズ・ネオアブソリュート} \\
Rose petal & \mbox{ローズペタル};\allowbreak\ \mbox{ローズ・ペタル} \\
Rosemary & \mbox{ローズマリー};\allowbreak\ \mbox{チュニジア産ローズマリー};\allowbreak\ \mbox{モロッコ産ローズマリー} \\
Rose leaf & \mbox{ローズ・リーフ} \\
Lotus & \mbox{ロータス};\allowbreak\ \mbox{ロータス（蓮）} \\
Lotus flower & \mbox{ロータスフラワー};\allowbreak\ \mbox{ロータス・フラワー} \\
Wyoming cotton flower & \mbox{ワイオミング・コットン・フラワー} \\
Wild Achillea & \mbox{ワイルド・アキレア} \\
Wild orchid & \mbox{ワイルド・オーキッド} \\
Wild strawberry & \mbox{ワイルドストロベリー};\allowbreak\ \mbox{ワイルド・ストロベリー} \\
Wacapou & \mbox{ワカポー（木材）} \\
Russian leather & \mbox{ロシアンレザー};\allowbreak\ \mbox{ロシアン・レザー} \\
Bulgarian red rose & \mbox{ブルガリアン・レッドローズ} \\
Violet leaf absolute & \mbox{ヴァイオレットリーフ・アブソリュート} \\
Violet leaf absolute & \mbox{ヴァイオレットリーフ・アブソルート} \\
Violet leaves & \mbox{ヴァイオレット・リーブス} \\
Vanity rose & \mbox{ヴァニティ・ローズ} \\
Rooibos & \mbox{南アフリカ原産ルイボス} \\
Japanese pear & \mbox{和ナシ};\allowbreak\ \mbox{和梨} \\
Citrus fruits & \mbox{柑橘系};\allowbreak\ \mbox{柑橘類} \\
Yuzu & \mbox{柚子};\allowbreak\ \mbox{日本産柚子} \\
Daidai & \mbox{橙（ダイダイ）} \\
Grassland & \mbox{草原（イネ科）} \\
\end{longtable}
\endgroup

\twocolumngrid

\endgroup
\color{black}
\clearpage
\let\PREpendingbibfont\bibfont
\renewcommand{\bibfont}{\PREpendingbibfont\color{black}}
\bibliography{references}

@article{Jaynes_1957,
  title = {Information Theory and Statistical Mechanics},
  volume = {106},
  url = {http://dx.doi.org/10.1103/PhysRev.106.620},
  doi = {10.1103/physrev.106.620},
  number = {4},
  journal = {Physical Review},
  author = {Jaynes, E. T.},
  year = {1957},
  pages = {620--630},
}

@article{Schneidman_2006,
  title = {Weak pairwise correlations imply strongly correlated network states in a neural population},
  volume = {440},
  url = {http://dx.doi.org/10.1038/nature04701},
  doi = {10.1038/nature04701},
  number = {7087},
  journal = {Nature},
  author = {Schneidman, Elad and Berry, II, Michael J. and Segev, Ronen and Bialek, William},
  year = {2006},
  pages = {1007--1012},
}

@article{Roudi_2009,
  title = {Pairwise Maximum Entropy Models for Studying Large Biological Systems: When They Can Work and When They Can’t},
  volume = {5},
  url = {http://dx.doi.org/10.1371/journal.pcbi.1000380},
  doi = {10.1371/journal.pcbi.1000380},
  number = {5},
  journal = {PLoS Computational Biology},
  author = {Roudi, Yasser and Nirenberg, Sheila and Latham, Peter E.},
  year = {2009},
  pages = {e1000380},
}

@article{Schneidman_2003,
  title = {Network Information and Connected Correlations},
  volume = {91},
  url = {http://dx.doi.org/10.1103/PhysRevLett.91.238701},
  doi = {10.1103/physrevlett.91.238701},
  number = {23},
  journal = {Physical Review Letters},
  author = {Schneidman, Elad and Still, Susanne and Berry, II, Michael J. and Bialek, William},
  year = {2003},
  pages = {238701},
}

@misc{CahiersDeMode_PerfumeEncyclopedia,
  author = {{Cahiers de Mode}},
  note = {\url{https://cahiersdemode.com/}},
}

@article{Cressie_1984,
  title = {Multinomial Goodness-Of-Fit Tests},
  volume = {46},
  url = {http://dx.doi.org/10.1111/j.2517-6161.1984.tb01318.x},
  doi = {10.1111/j.2517-6161.1984.tb01318.x},
  number = {3},
  journal = {Journal of the Royal Statistical Society Series B: Statistical Methodology},
  author = {Cressie, Noel and Read, Timothy R.C.},
  year = {1984},
  pages = {440--464},
}

@article{Kubo_1957,
  title = {Statistical-Mechanical Theory of Irreversible Processes. {I}. {General} Theory and Simple Applications to Magnetic and Conduction Problems},
  volume = {12},
  url = {http://dx.doi.org/10.1143/JPSJ.12.570},
  doi = {10.1143/jpsj.12.570},
  number = {6},
  journal = {Journal of the Physical Society of Japan},
  author = {Kubo, Ryogo},
  year = {1957},
  pages = {570--586},
}

@article{Kappen_1998,
  title = {Efficient Learning in {Boltzmann} Machines Using Linear Response Theory},
  volume = {10},
  url = {http://dx.doi.org/10.1162/089976698300017386},
  doi = {10.1162/089976698300017386},
  number = {5},
  journal = {Neural Computation},
  author = {Kappen, H. J. and Rodr{\'i}guez, F. B.},
  year = {1998},
  pages = {1137--1156},
}

@article{Gelfand_1990,
  title = {Sampling-Based Approaches to Calculating Marginal Densities},
  volume = {85},
  url = {http://dx.doi.org/10.1080/01621459.1990.10476213},
  doi = {10.1080/01621459.1990.10476213},
  number = {410},
  journal = {Journal of the American Statistical Association},
  author = {Gelfand, Alan E. and Smith, Adrian F. M.},
  year = {1990},
  pages = {398--409},
}

@inproceedings{kingma2015adammethodstochasticoptimization,
  title = {{Adam}: A Method for Stochastic Optimization},
  author = {Kingma, Diederik P. and Ba, Jimmy Lei},
  year = {2015},
  eprint = {1412.6980},
  archiveprefix = {arXiv},
  primaryclass = {cs.LG},
  url = {https://arxiv.org/abs/1412.6980v5},
  booktitle = {International Conference on Learning Representations},
}

@article{Battiston_2020,
  title = {Networks beyond pairwise interactions: Structure and dynamics},
  volume = {874},
  url = {http://dx.doi.org/10.1016/j.physrep.2020.05.004},
  doi = {10.1016/j.physrep.2020.05.004},
  journal = {Physics Reports},
  author = {Battiston, Federico and Cencetti, Giulia and Iacopini, Iacopo and Latora, Vito and Lucas, Maxime and Patania, Alice and Young, Jean-Gabriel and Petri, Giovanni},
  year = {2020},
  pages = {1--92},
}

@article{Ganmor_2011_PNAS,
  title = {Sparse low-order interaction network underlies a highly correlated and learnable neural population code},
  volume = {108},
  url = {http://dx.doi.org/10.1073/pnas.1019641108},
  doi = {10.1073/pnas.1019641108},
  number = {23},
  journal = {Proceedings of the National Academy of Sciences},
  author = {Ganmor, Elad and Segev, Ronen and Schneidman, Elad},
  year = {2011},
  pages = {9679--9684},
}

@article{Tka_ik_2014,
  title = {Searching for Collective Behavior in a Large Network of Sensory Neurons},
  volume = {10},
  url = {http://dx.doi.org/10.1371/journal.pcbi.1003408},
  doi = {10.1371/journal.pcbi.1003408},
  number = {1},
  journal = {PLoS Computational Biology},
  author = {Tka{\v{c}}ik, Ga{\v{s}}per and Marre, Olivier and Amodei, Dario and Schneidman, Elad and Bialek, William and Berry, II, Michael J.},
  year = {2014},
  pages = {e1003408},
}

@article{Shimazaki_2015,
  title = {Simultaneous silence organizes structured higher-order interactions in neural populations},
  volume = {5},
  url = {http://dx.doi.org/10.1038/srep09821},
  doi = {10.1038/srep09821},
  number = {1},
  journal = {Scientific Reports},
  author = {Shimazaki, Hideaki and Sadeghi, Kolia and Ishikawa, Tomoe and Ikegaya, Yuji and Toyoizumi, Taro},
  year = {2015},
  pages = {9821},
}

@article{Stephens_2010,
  title = {Statistical mechanics of letters in words},
  volume = {81},
  url = {http://dx.doi.org/10.1103/PhysRevE.81.066119},
  doi = {10.1103/physreve.81.066119},
  number = {6},
  journal = {Physical Review E},
  author = {Stephens, Greg J. and Bialek, William},
  year = {2010},
  pages = {066119},
}

@article{Kinouchi_2008,
  title = {The non-equilibrium nature of culinary evolution},
  volume = {10},
  url = {http://dx.doi.org/10.1088/1367-2630/10/7/073020},
  doi = {10.1088/1367-2630/10/7/073020},
  number = {7},
  journal = {New Journal of Physics},
  author = {Kinouchi, Osame and Diez-Garcia, Rosa W and Holanda, Adriano J and Zambianchi, Pedro and Roque, Antonio C},
  year = {2008},
  pages = {073020},
}

@article{Ahn_2011,
  title = {Flavor network and the principles of food pairing},
  volume = {1},
  url = {http://dx.doi.org/10.1038/srep00196},
  doi = {10.1038/srep00196},
  number = {1},
  journal = {Scientific Reports},
  author = {Ahn, Yong-Yeol and Ahnert, Sebastian E. and Bagrow, James P. and Barab{\'a}si, Albert-L{\'a}szl{\'o}},
  year = {2011},
  pages = {196},
}

@article{Caprioli_2025,
  title = {The networks of ingredient combinations as culinary fingerprints of world cuisines},
  volume = {9},
  url = {http://dx.doi.org/10.1038/s41538-025-00588-4},
  doi = {10.1038/s41538-025-00588-4},
  number = {1},
  journal = {npj Science of Food},
  author = {Caprioli, Claudio and Kulkarni, Saumitra and Battiston, Federico and Iacopini, Iacopo and Santoro, Andrea and Latora, Vito},
  year = {2025},
  pages = {242},
}

@misc{satarifard2026culturalevolutionperfumes1900,
  title = {Cultural Evolution of Perfumes since 1900},
  author = {Vahid Satarifard and Fabian Baumann and Geetanjali Minsky and Laura Sisson and Lou M. Haux and Christophe Laudamiel and Nicholas A. Christakis},
  year = {2026},
  eprint = {2607.19520},
  archiveprefix = {arXiv},
  primaryclass = {physics.soc-ph},
  url = {https://arxiv.org/abs/2607.19520v1},
}

@article{Amari_2001,
  title = {Information geometry on hierarchy of probability distributions},
  volume = {47},
  url = {http://dx.doi.org/10.1109/18.930911},
  doi = {10.1109/18.930911},
  number = {5},
  journal = {IEEE Transactions on Information Theory},
  author = {Amari, S.-I.},
  year = {2001},
  pages = {1701--1711},
}

@article{Geman_1984,
  title = {Stochastic Relaxation, {Gibbs} Distributions, and the {Bayesian} Restoration of Images},
  author = {Geman, Stuart and Geman, Donald},
  journal = {IEEE Transactions on Pattern Analysis and Machine Intelligence},
  volume = {PAMI-6},
  number = {6},
  pages = {721--741},
  year = {1984},
  doi = {10.1109/TPAMI.1984.4767596},
}

@article{Ekeberg_2013,
  title = {Improved Contact Prediction in Proteins: Using Pseudolikelihoods to Infer {Potts} Models},
  author = {Ekeberg, Magnus and L{\"o}vkvist, Cecilia and Lan, Yueheng and Weigt, Martin and Aurell, Erik},
  journal = {Physical Review E},
  volume = {87},
  pages = {012707},
  year = {2013},
  doi = {10.1103/PhysRevE.87.012707},
}

@article{Montemurro_2011,
  author = {Montemurro, Marcelo A. and Zanette, Dami{\'{a}}n H.},
  title = {Universal Entropy of Word Ordering Across Linguistic Families},
  journal = {PLOS ONE},
  volume = {6},
  number = {5},
  pages = {e19875},
  year = {2011},
  doi = {10.1371/journal.pone.0019875},
  url = {https://doi.org/10.1371/journal.pone.0019875}
}

@article{Herrera_2021,
  author = {Herrera, Juan C. S.},
  title = {The contribution of network science to the study of food recipes. {A} review paper},
  journal = {Appetite},
  volume = {159},
  pages = {105048},
  year = {2021},
  doi = {10.1016/j.appet.2020.105048},
  url = {https://doi.org/10.1016/j.appet.2020.105048}
}

@article{Chastrette_1988,
  author = {Chastrette, Maurice and Elmouaffek, Allal and Sauvegrain, Philippe},
  title = {A multidimensional statistical study of similarities between 74 notes used in perfumery},
  journal = {Chemical Senses},
  volume = {13},
  number = {2},
  pages = {295--305},
  year = {1988},
  doi = {10.1093/chemse/13.2.295},
  url = {https://doi.org/10.1093/chemse/13.2.295}
}

@article{Cai_2024,
  author = {Cai, Jiaxiao and Tuo, Suxing and Li, Yanchun and Lu, Hongbing and Wu, Yizi and Zou, You and Ma, Zhen and Cui, Yuqi and Kong, Bo and Zhong, Kejun},
  title = {A rapid approach based on empirical formulas and graph traversal algorithms to automatic generation of fragrance formulas},
  journal = {Heliyon},
  volume = {10},
  number = {24},
  pages = {e40873},
  year = {2024},
  doi = {10.1016/j.heliyon.2024.e40873},
  url = {https://doi.org/10.1016/j.heliyon.2024.e40873}
}

@article{Morcos_2011,
  author = {Morcos, Faruck and Pagnani, Andrea and Lunt, Bryan and Bertolino, Arianna and Marks, Debora S. and Sander, Chris and Zecchina, Riccardo and Onuchic, Jos{\'{e}} N. and Hwa, Terence and Weigt, Martin},
  title = {Direct-coupling analysis of residue coevolution captures native contacts across many protein families},
  journal = {Proceedings of the National Academy of Sciences},
  volume = {108},
  number = {49},
  pages = {E1293--E1301},
  year = {2011},
  doi = {10.1073/pnas.1111471108},
  url = {https://doi.org/10.1073/pnas.1111471108}
}

@article{Lambiotte_2019,
  author = {Lambiotte, Renaud and Rosvall, Martin and Scholtes, Ingo},
  title = {From networks to optimal higher-order models of complex systems},
  journal = {Nature Physics},
  volume = {15},
  number = {4},
  pages = {313--320},
  year = {2019},
  doi = {10.1038/s41567-019-0459-y},
  url = {https://doi.org/10.1038/s41567-019-0459-y}
}

@article{Casella_Robert_1996,
  author = {Casella, George and Robert, Christian P.},
  title = {{Rao-Blackwellisation} of sampling schemes},
  journal = {Biometrika},
  volume = {83},
  number = {1},
  pages = {81--94},
  year = {1996},
  doi = {10.1093/biomet/83.1.81},
  url = {https://doi.org/10.1093/biomet/83.1.81}
}

@article{Chodrow_2020,
  author = {Chodrow, Philip S.},
  title = {Configuration models of random hypergraphs},
  journal = {Journal of Complex Networks},
  volume = {8},
  number = {3},
  pages = {cnaa018},
  year = {2020},
  doi = {10.1093/comnet/cnaa018},
  url = {https://doi.org/10.1093/comnet/cnaa018}
}

@article{Kubo_1966,
  author = {Kubo, Ryogo},
  title = {The fluctuation-dissipation theorem},
  journal = {Reports on Progress in Physics},
  volume = {29},
  number = {1},
  pages = {255--284},
  year = {1966},
  doi = {10.1088/0034-4885/29/1/306},
  url = {https://doi.org/10.1088/0034-4885/29/1/306}
}

@article{Victor_2013,
  author  = {Victor, Jonathan D. and Thengone, Daniel J. and Conte, Mary M.},
  title   = {Perception of second- and third-order orientation signals and their interactions},
  journal = {Journal of Vision},
  volume  = {13},
  number  = {4},
  pages   = {21},
  year    = {2013},
  doi     = {10.1167/13.4.21}
}
\end{document}